\documentclass[12pt,english]{revtex4-1}
\usepackage[T1]{fontenc}
\usepackage[latin9]{inputenc}
\usepackage{geometry}
\usepackage{pdfcolmk}
\usepackage{float}
\usepackage{amsmath}
\usepackage{amssymb}
\usepackage{graphicx}
\PassOptionsToPackage{normalem}{ulem}
\usepackage{ulem}



\usepackage{babel}
\begin{document}

\title{Partition-free approach to open quantum systems in harmonic environments: an exact stochastic Liouville equation}

\author{G.M.G. McCaul, C.D. Lorenz and L. Kantorovich}

\address{Physics Department, King's College London, The Strand, London, WC2R
2LS, United Kingdom}
\begin{abstract}We present a partition-free approach to the evolution of density matrices for 
open quantum systems coupled to a harmonic environment. The 
influence functional formalism combined with a two-time Hubbard-Stratonovich 
transformation allows us to derive a set of exact differential equations 
for the reduced density matrix of an open system, termed the Extended Stochastic 
Liouville-von Neumann equation. Our approach generalises previous work 
based on Caldeira-Leggett models and  a partitioned initial density matrix. This provides a simple, 
yet exact, closed-form description for the evolution of open systems from 
equilibriated initial conditions. The applicability of this model and 
the potential for numerical implementations are also discussed.
\end{abstract}
\maketitle

\section{Introduction}

Much of the work in the canon of physics has been derived under an
assumption of isolation, where the system of interest has no interaction
with its environment. Often, particularly in the classical regime,
this approximation has been successful in generating accurate predictions.
There are however numerous systems whose behaviour cannot be explained
by their actions in a vacuum \cite{Ford2015}. In these cases stochastic
terms are used, often as an \emph{a priori} part of the model (and
without proper justification), to capture the effect of the environment.
Brownian motion is the most famous case of this technique in classical
physics, but quantum physics and its applications have many examples
where a similarly careful treatment of external effects is required
\cite{Weiss-1999,50-years-Kramer-1990,Melnikov1991}. These systems
can collectively be termed \emph{open dissipative quantum systems},
and the problem of how to most accurately model them remains an active
field of research. 

Approaches to these systems can be split into two broad categories.
The first method uses the paradigmatic example of a damped system,
where the damping is an effective loss-mechanism that approximates
the environment's effect and fluctuations are neglected. A typical
example of this is the early work of Kerner and Stevens on sets of
damped harmonic oscillators \cite{Stevens1958,Kerner1958}. The basis
of this method in classical, phenomological equations means that it
is capable of providing exact solutions for some simple systems, such
as the damped harmonic oscillator. These solutions are however undermined
by being fundamentally incompatible with quantum mechanics. This can
be illustrated by the fact that there are no \emph{time-independent}
Hamiltonians that can replicate the equation of motion for a damped
oscillator, 

\begin{equation}
m\ddot{x}+\alpha\dot{x}+m\omega^{2}x=0
\end{equation}
which has frequency $\omega$ and friction $\alpha$. While there
exists a \emph{time-dependent} Hamiltonian that leads to this equation
of motion \cite{Kanai01121948}, after quantisation the fundamental
commutation relation becomes time-dependent \cite{Senitzky1960}.
This unphysical result means that another approach to dissipative
systems, to be detailed below, is the method of choice. 

In this approach, pioneered by Callen, Welton, Senitzky and Lax, dissipative
systems are modelled as a primary system (the ``open system'') of
interest coupled to an explicit secondary system (the ``environment''
or ``heat bath'') which together describe the overall system being
modeled (the ``total system'') \cite{Callen1951,Lax1963,Senitzky1960}.
In comparison to the first method, this model is lossless when considering
the total system, and incorporates both the dissipation \emph{and
}fluctuations experienced by the open system as a consequence of its
explicit coupling to the environment. Combining this model with appropriate
approximations (e.g. weak coupling between the open system and environment)
allows quantum master equations to be derived, which retain the correct
behaviour in the classical limit \cite{Ford1965,Benguria1981,Schmid1982,Lindenberg1984,Cortes1985}.

The general scheme then is to treat the coupled systems as a single
closed sytem which can be straightforwardly quantised. The environmental
coordinates can then be eliminated in order to obtain an equation
of motion for the primary system. In practice the functional form
of the environment (secondary system) and its coupling must be chosen
subject to several conditions. For example in the high-temperature
classical limit we expect to recover a classical Brownian motion.
In addition, if the summation over environmental coordinates is to
be exact, yet analytically tractable, the choice of environment is
largely restricted to a set of harmonic oscillators, with a bilinear
coupling to the open system. A particularly popular model is the Caldeira-Leggett
(CL) Hamiltonian \cite{Caldeira1983}:

\begin{equation}
H=H_{q}(q)+\frac{1}{2}\sum_{i}\left(m_{i}\dot{x}_{i}^{2}+m_{i}\omega_{i}^{2}x_{i}^{2}\right)-q\sum_{i}c_{i}x_{i}+\frac{q^{2}}{2}\sum_{i}\frac{c_{i}^{2}}{m_{i}\omega_{i}^{2}}\label{eq:Caldeira_Legget_Model}
\end{equation}
This model couples the open system (described by the coordinate $q)$
to an environment of independent harmonic oscillators (masses $m_{i}$,
frequencies $\omega_{i}$, and displacement coordinates $x_{i}$)
with each oscillator being coupled to the open system with a strength
$c_{i}$. The final term is a counter-term included to enforce translational
invariance on the system and eliminate quasi-static effects \cite{PhysRevA.61.022107counterterm}. 

Recently, a more general Hamiltonian of the combined system (the open
system and harmonic environment) was introduced \cite{my-SBC-1} which
is only linear with respect to the environmental variables, but remains
arbitrary with respect to the positions of atoms in the open system
(this model is detailed in section \ref{sec:Model}). In this Hamiltonian
interactions within the environment are not diagonalised. This is
convenient because all parameters of the environment and its interaction
with the open system can then be extracted by expanding the Hamiltonian
of the combined system in atomic displacements in the bath and keeping
only harmonic terms, i.e. the open system can be considered as a part
of the expansion of the total system. This rather general choice of
total system Hamiltonian enables one to derive classical equations
of motion {[}in the form of the Generalised Langevin Equation (GLE){]}
for the atoms in the open system \cite{my-SBC-1} and propose an efficient
numerical scheme for solving them \cite{Lorenzo-GLE-2014,Herve-GLE-2015,Herve-GLE-2016}.
This method has been recently generalised to the fully quantum case
\cite{QGLE-2016} where it was shown, using the method based on directly
solving the Liouville equation, that equations of motion for the observable
positions of atoms in the open system have the GLE form with friction
memory and non-Gaussian random force terms. Although this method enables
one to develop the general structure of the equations to be expected
for the open system, this method lacks an exact mechanism for establishing
the necessary expressions for the random force correlation functions. 

In the study of quantum Brownian motion, the path integral representation
has been perhaps the most fruitful. Some specific successful applications
include tunnelling and decay rate calculations (Kramer's problem)
\cite{Hanggi1987,Grabert1987,Melnikov1991,Pollak1989,Affleck1981,50-years-Kramer-1990}
as well as recent first-principle derivations for the rate of processes
in instanton theory \cite{Richardson2015a,Richardson2015}. In particular
the Feynman-Vernon influence functional formalism \cite{Feynman-Vernon-1963}
can be used to exactly calculate the effect of the environment on
the open system using path integrals. Approximations such as weak
coupling between the primary system and environment are no longer
necessary. Path integrals also remove the need for an explicit quantisation
of the system Hamiltonian, as in this formalism quantum-mechanical
propagators are represented as phase-weighted sums over trajectories,
where the phase associated to each trajectory is proportional to the
action of that path in the classical system \cite{FeynmanHibbs}.
A useful consequence of this is that the classical limit is easily
obtained \cite{TechniquesApplicationsPathIntegration}, and the quantisation
of the system is automatic when choosing this representation. Finally,
and probably most importantly, bath degrees of freedom can be integrated
out exactly if the environment is harmonic and interacts with the
open system via an expression that is at most up to the second order
in its displacements. 

The key simplification of the Feynman-Vernon approach is that initially
the density matrix of the total system $\hat{\rho}_{0}^{\mbox{tot}}$
can be partitioned, 

\begin{equation}
\hat{\rho}_{0}^{\mbox{tot}}=\hat{\rho}_{0}\otimes\hat{\rho}_{0}^{X}\label{eq:FV_Direct_Prod}
\end{equation}
i.e. it can be expressed as a direct product of the initial density
matrices of the open system $\hat{\rho}_{0}$ and the environment
$\hat{\rho}_{0}^{X}$, where each subsystem has equilibriated separately. 

In the context of open, dissipative quantum systems, much work has
been done using this formalism, expanding the methodology of the Feynman-Vernon
influence functional for both exact and approximate results \cite{Smith1987,Makri1989,Allinger1989}.
Using this model, quantum Langevin equations for the reduced density
matrix have been rigorously derived using path integrals \cite{Caldeira1983,Ford-Kac-JST-1987,Gardiner-1988,Sebastian1981,Leggett1987,van_Kampen-1997}.
In special cases, further analytical results have also been obtained
by Kleinert \cite{Kleinert1995,Kleinertbook} and Tsusaka \cite{Tsusaka1999}.
Generalisations of these results to anharmonic baths produce approximate
but more realistic models \cite{Bhadra2016,McDowell2000}, while time-dependent
heat exchange can also be exactly included \cite{Carrega2015}. Parallel
to this is the work of Stockburger, exactly deriving a stochastic
Liouville-von Neumann (SLN) equation, and applying it to two-level
systems \cite{Stockburger2004}. Approaches based on influence functionals
have also found use in the real time numerical simulations of dissipative
systems \cite{Banerjee2015,Makri2014,Makri1998,Dattani2012,Habershon2013,Herrero2014,Wang2007}.
With this corpus of techniques, path integrals (and specifically influence
functionals) represent a powerful and flexible formalism that can
be used to attack the problem of open quantum systems. 

So far, we have been discussing methods based on initially partitioning
the total system. The initial condition of Eq. (\ref{eq:FV_Direct_Prod})
is however unphysical, as it is impossible in a real experiment to
``prepare'' a quantum system with the interaction between the open
system and environment switched off, prior to any perturbation being
applied. As a result, the transient behaviour we predict for perturbations
away from a partitioned initial condition will always be spurious
due to the artificial equilibriation of each system seperately. If
we wish to extract the exact transient dynamics of an open system
we must therefore use a more realistic, non-partitioned initial condition. 

Fortunately, the influence functional formalism has the capacity to
naturally generalise the initial conditions of the overall system
and environment, rendering the assumption of a partitioned initial
state unnecessary. This possibility was first noted by Smith and Caldeira
\cite{Smith1987}, before being properly explored by Grabert, Ingold
and Schramm \cite{Grabert1988}, who derived the time dependent expression
for the reduced density matrix of an open system where all path integrals
associated with the environment are fully eliminated. In this partition-free
case, the limits on our ability to describe the reduced dynamics via
a Liouville operator have been derived by Karrlein and Grabert \cite{Karrlein1997}.
In this work however, no differential equation for the reduced density
matrix was derived, and the authors still used a simplified CL Hamiltonian.
We also note that a differential equation for the \emph{equilibrium}
reduced density matrix for the CL Hamiltonian was obtained using path
integrals in Ref. \cite{Moix2012} and is consistent with our results.

In this paper, we derive, using the path integral formalism, a set
of stochastic differential equations for the reduced density matrix
of an open system which describe its dynamics \emph{exactly}. The
derived equation does not have the GLE form obtained previously in
Ref. \cite{QGLE-2016}. Indeed, it does not have a clearly defined
friction term and the stochastic fields it contains are Gaussian.
Nevertheless, our Hamiltonian is identical to the one used in Ref.
\cite{QGLE-2016}, which is more general than the CL Hamiltonian.
Using it, we obtain a system of first order stochastic differential
equations over real and imaginary time that exactly describe the evolution
of the state of a dissipative quantum system for partition-free initial
conditions. These equations, which we term the Extended Stochastic
Liouville Equation (ESLN), represent both a synthesis and extension
of the work outlined above, allowing for a simple and exact closed
form description of an arbitrary open system evolving from realistic
initial conditions. The derivation of the ESLN, (and therefore the
paper itself) will be organised as follows: 

Section \ref{sec:Model} details the model employed, and the class
of applicable initial conditions. In section \ref{sec:Influence_Functional}
the path integral representation for the density matrix of the primary
system will be introduced, along with the influence functional and
its explicit evaluation. In section \ref{sec:The-Two-Time-Hubbard-Stratonovich}
the two-time Hubbard-Stratonovich transformation is applied to the
influence functional found in the previous section, introducing the
corresponding complex Gaussian stochastic fields. Section \ref{sec:The-Extended-Stochastics}
presents the path integral describing the reduced density matrix of
the primary system and the operator ESLN equations of motion that
it implies, which represents the central result of this work. These
equations account for both the generalised Hamiltonian and partition-free
initial conditions. Finally, section \ref{sec:Discussion} concludes
the paper with a discussion of the ESLN, its connection to previous
results and the potential for numerical implementations.

\section{Model\label{sec:Model}}

\begin{figure}[H]
\begin{centering}
\includegraphics[height=9cm]{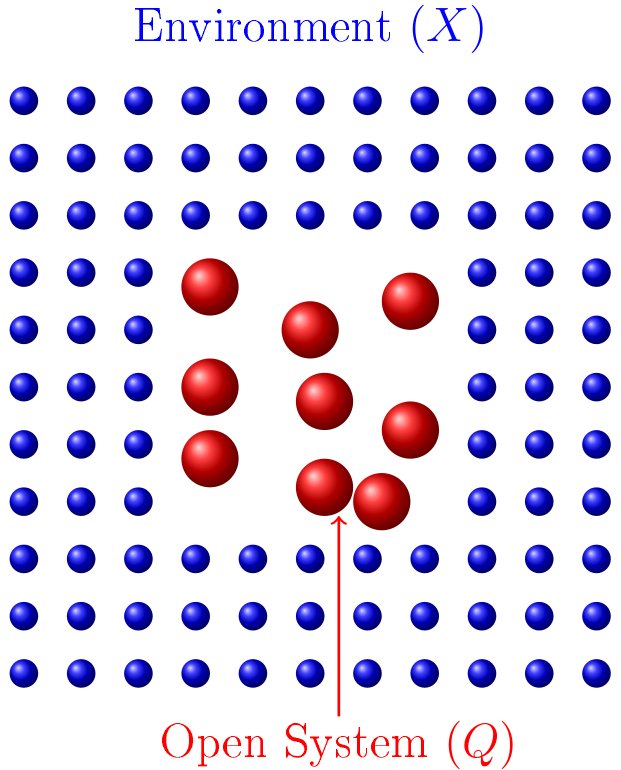}
\par\end{centering}

\caption{Schematic of the system. The $Q$ system will be described by the
$q$ coordinates, and its environment, the $X$ system, with $\xi$
coordinates ($x$ in normal modes).\label{fig:Schematic}}
\end{figure}

Consider a many-body phonon system of the type shown in Figure \ref{fig:Schematic}.
It consists of a general central system (the open system), described
by coordinates $q$, acting under an arbitrary Hamiltonian $H_{q}\left(q\right)$.
The secondary system (the environment) is composed of $M$ harmonic
oscillators (with masses $m_{i}$) coupled both internally and with
the open system. The open system may be subjected to time-dependent
external fields. The environment uses displacement coordinates $\xi_{i}$
and the interaction between the two systems is linear in $\xi\equiv\left\{ \xi_{i}\right\} $
but arbitrary in $q$:

\begin{equation}
H_{\textrm{tot}}(q,\xi)=H_{q}(q)+\frac{1}{2}\sum_{i=1}^{M}m_{i}\dot{\xi}_{i}^{2}+\frac{1}{2}\sum_{i,j=1}^{M}\varLambda_{ij}\xi_{i}\xi_{j}-\sum_{i}^{M}f_{i}(q)\xi_{i}\label{eq:Model_Hamiltonian}
\end{equation}
This Hamiltonian differs from the standard CL Hamiltonian in Eq. (\ref{eq:Caldeira_Legget_Model})
in two important respects. First, the interaction between the primary
and secondary systems is no longer strictly bilinear, but can depend
arbitrarily on $q$. In addition, the atomic displacements that form
the environment are now coupled to each other as well as the system,
with the coupling described by the force-constant matrix $\varLambda_{ij}$.
These alterations will have a material effect on our results. We also
note the counter-term found in Eq. (\ref{eq:Caldeira_Legget_Model})
has been dropped as it is no longer needed, since when the Hamiltonian
of an arbitrary combined system is expanded in the power series in
terms of atomic displacements $\xi_{i}$, this kind of term does not
appear. In this sense our model Hamiltonian is the second-order expansion
of \emph{any} conceivable system-bath Hamiltonian. 

The density matrix evolves in the usual manner according to the Liouville
equation:

\begin{equation}
\hat{\rho}^{\textrm{tot}}(t)=\widehat{U}(t;t_{0})\hat{\rho}^{\textrm{tot}}(t_{0})\widehat{U}^{\dagger}(t;t_{0})\label{eq:Solution_of_Liouville_Equation}
\end{equation}
where 

\begin{equation}
\widehat{U}(t;t_{0})=\exp\left[-\frac{i}{\hbar}\int_{t_{0}}^{t}\mathrm{d}t^{\prime}\ \widehat{H}_{\textrm{tot}}(t^{\prime})\right]\label{eq:Operator_Propagator_Definition}
\end{equation}
is the corresponding evolution operator. Importantly we need not assume
that the system Hamiltonian $H_{q}(q)$ is time-independent. i.e.
$H_{q}\left(q\right)\equiv H_{q}\left(q,t\right)$. The dynamics of
the open system are found by tracing the full density matrix over
the $\xi$ coordinates:

\begin{equation}
\hat{\rho}(t)=\textrm{Tr}_{\xi}\left[\hat{\rho}^{\textrm{tot}}(t)\right]
\end{equation}
while the total and reduced density matrices in coordinate space are,
respectively:

\begin{equation}
\rho_{t}^{\textrm{tot}}\left(q,\xi;q^{\prime},\xi^{\prime}\right)=\left\langle q,\xi\right|\hat{\rho}^{\textrm{tot}}(t)\left|q^{\prime},\xi^{\prime}\right\rangle 
\end{equation}

\begin{equation}
\rho_{t}(q,q^{\prime})=\left\langle q\right|\hat{\rho}(t)\left|q^{\prime}\right\rangle 
\end{equation}
The propagators in this space are given by:

\begin{equation}
U\left(q,\xi,t;\bar{q},\bar{\xi},t_{0}\right)=\left\langle q,\xi\left|\widehat{U}\left(t;t_{0}\right)\right|\bar{q},\bar{\xi}\right\rangle \label{eq:U_in_Coordinates}
\end{equation}
\begin{equation}
\left\langle \bar{q},\bar{\xi}\left|\widehat{U}^{\dagger}\left(t;t_{0}\right)\right|q,\xi\right\rangle =\left\langle \bar{q},\bar{\xi}\left|\widehat{U}\left(t_{0};t\right)\right|q,\xi\right\rangle =U\left(\bar{q},\bar{\xi},t_{0};q,\xi,t\right)\label{eq:U_Dagger_in_Coordinates}
\end{equation}
The second equality has been constructed to demonstrate that in coordinates,
$U^{\dagger}$ has the form of a backward propagation in time. Setting
$t_{0}=0$ for convenience, the open system density matrix in the
coordinate representation is:

\begin{equation}
\rho_{t}(q,q')=\int\mathrm{d}\overline{\xi}\ \mathrm{d}\overline{\xi}^{\prime}\ \mathrm{d}\bar{q}\ \mathrm{d}\bar{q}^{\prime}\ \mathrm{d}\xi\mathrm{\ d}\xi^{\prime}\ \delta\left(\xi-\xi^{\prime}\right)U(q,\xi,t;\bar{q},\bar{\xi},0)\rho_{0}^{\textrm{tot}}(\bar{q,}\bar{\xi};\bar{q}^{\prime},\bar{\xi}^{\prime})U(\bar{q}^{\prime},\overline{\xi}^{\prime},0;q^{\prime},\xi^{\prime},t)
\end{equation}

At this point we transform to a normal mode representation $\xi\rightarrow x=\left\{ x_{\lambda}\right\} $,
where
\[
x_{\lambda}=\sum_{i}^{M}\sqrt{m_{i}}e_{\lambda i}\xi_{i}\ \ ,\ \ \ \xi_{i}=\sum_{\lambda}^{M}\frac{1}{\sqrt{m_{i}}}e_{i\lambda}x_{\lambda}
\]
and $e_{\lambda}=\left\{ e_{\lambda i}\right\} $ are eigenvectors
of the dynamical matrix $D=\left\{ D_{ij}\right\} $, where $D_{ij}=\Lambda_{ij}/\sqrt{m_{i}m_{j}}$,
with eigenvalues $\omega_{\lambda}^{2}$. The eigenvectors satisfy
the usual orthogonality, $e_{\lambda}^{T}e_{\lambda^{\prime}}=\delta_{\lambda\lambda^{\prime}}$,
and completeness, $\sum_{\lambda}e_{\lambda}e_{\lambda}^{T}=1$, conditions
(the superscript $T$ stands for transpose). Applying these transformations,
the Hamiltonian can be expressed as:

\begin{equation}
H_{\textrm{tot}}(q,x)=H_{q}(q)+\frac{1}{2}\sum_{\lambda=1}^{M}\left(\dot{x}_{\lambda}^{2}+\omega_{\lambda}^{2}x_{\lambda}^{2}\right)-\sum_{\lambda}g_{\lambda}(q)x_{\lambda}\label{eq:Normal_Mode_Transformed_Hamiltonian}
\end{equation}
where 
\begin{equation}
g_{\lambda}(q)=\sum_{i}^{M}\frac{1}{\sqrt{m_{i}}}e_{\lambda i}f_{i}(q)\;,\quad f_{i}(q)=\sqrt{m_{i}}\sum_{\lambda}^{M}e_{i\lambda}g_{\lambda}(q)\label{eq:Normal_Mode_Interaction_Term}
\end{equation}
The reduced density matrix is now given by: 

\begin{equation}
\rho_{t}(q,q^{\prime})=\int\mathrm{d}\bar{x}\mathrm{d}\bar{x}^{\prime}\mathrm{d}x\ \mathrm{d}\bar{q}\mathrm{d}\bar{q}^{\prime}\,U\left(q,x,t;\bar{q},\bar{x},0\right)\rho_{0}^{\textrm{tot}}\left(\bar{q,}\bar{x};\bar{q}^{\prime},\bar{x}^{\prime}\right)U\left(\bar{q}^{\prime},\bar{x}^{\prime},0;q^{\prime},x,t\right)\label{eq:Reduced_Density_Normal_Mode_Coordinates}
\end{equation}
Before Eq. (\ref{eq:Reduced_Density_Normal_Mode_Coordinates}) can
be solved, we must specify the form of the initial density matrix
$\rho_{0}^{\textrm{tot}}$. As was explained in the Introduction,
in most systems of interest the interaction between the primary system
and its environment is an integral part of the system and hence one
cannot assume the two systems are initially partitioned. One solution
employed by Grabert\emph{ et al.} \cite{Grabert1988} is to consider
the full interacting system as being allowed to equilibrate with some
time-independent Hamiltonian $H_{0}$ before applying any time-dependent
perturbation. In this case the initial state would then be described
by the canonical density matrix: 
\begin{equation}
\hat{\rho}_{0}^{\textrm{tot}}\equiv\hat{\rho}_{\beta}=\frac{1}{Z_{\beta}}\mathrm{e}^{-\beta H_{0}}\label{eq:Canonical_Density}
\end{equation}
where $\beta=1/k_{B}T$ is the inverse temperature and $Z_{\beta}=\mbox{Tr}\left(e^{-\beta H_{0}}\right)$
is the corresponding partition function of the entire system. Note
that a class of more general initial density matrices can be considered
\cite{Grabert1988}, however, here we shall limit ourselves only to
the canonical density matrix.

Having specified the initial conditions, the goal is now to derive
an equation of motion that will describe the \emph{exact} evolution
of the reduced density matrix $\rho_{t}\left(q,q^{\prime}\right)$
as given by Eq. (\ref{eq:Reduced_Density_Normal_Mode_Coordinates}).
To do this we will utilise the influence functional to eliminate the
environmental degrees of freedom in Eq. (\ref{eq:Reduced_Density_Normal_Mode_Coordinates}).

\section{The Path Integral Representation and Influence Functional \label{sec:Influence_Functional}}

To proceed we will insert the path integral representation of both
propagators and the initial density matrix into Eq. (\ref{eq:Reduced_Density_Normal_Mode_Coordinates}).
The expression for the forward propagator $U\left(q,x,t_{f};\bar{q},\bar{x},0\right)$
as a path integral up to a time $t_{f}$ is given by 
\begin{equation}
U\left(q,x,t_{f};\bar{q},\bar{x},0\right)=\int_{q\left(0\right)=\bar{q}}^{q\left(t_{f}\right)=q}\mathcal{D}q\left(t\right)\int_{x\left(0\right)=\bar{x}}^{x\left(t_{f}\right)=x}\mathcal{D}x\left(t\right)\,\exp\left(\frac{i}{\hbar}S\left[q\left(t\right),x\left(t\right)\right]\right)\label{eq:Propagator_Forward}
\end{equation}
with a similar definition for the backward propagator 
\begin{equation}
U\left(\bar{q}^{\prime},\bar{x}^{\prime},0;q^{\prime},x,t_{f}\right)=\int_{q^{\prime}\left(t_{f}\right)=q^{\prime}}^{q^{\prime}\left(0\right)=\overline{q}^{\prime}}\mathcal{D}q^{\prime}\left(t\right)\int_{x^{\prime}\left(t_{f}\right)=x}^{x^{\prime}\left(0\right)=\bar{x}^{\prime}}\mathcal{D}x^{\prime}\left(t\right)\,\exp\left(-\frac{i}{\hbar}S\left[q^{\prime}\left(t\right),x^{\prime}\left(t\right)\right]\right)\label{eq:Propagator_Backwards}
\end{equation}
The limits of the path integral in the second propagator are reversed
as compared to the first one to emphasize its backward nature, as
in Eq. (\ref{eq:U_Dagger_in_Coordinates}).

In both expressions the integration is performed with respect to both
the open system ($q,q^{\prime}$) and environment ($x,x^{\prime}$)
variables between the boundaries indicated. Here $S$ is the action
corresponding to the Hamiltonian in Eq. (\ref{eq:Normal_Mode_Transformed_Hamiltonian})
describing the total system. It is defined in both propagators in
the usual manner (i.e. the time integral of the Langrangian from $0$
to $t_{f}$), hence the extra negative in the exponent of the backwards
propagator. Integration over the environmental variables can be performed
exactly as the environment and interaction Hamiltonians added together
have the form of a set of displaced harmonic oscillators in the environment
variables. This means the path integral over environmental trajectories
is Gaussian, and can be evaluated (see, e.g., \cite{FeynmanHibbs,Feynman-Vernon-1963,Grabert1988}).
The propagator therefore becomes a path integral over the trajectories
of the open system only:
\begin{equation}
U\left(q,x,t_{f};\bar{q},\bar{x},0\right)=A\int_{q\left(0\right)=\bar{q}}^{q\left(t_{f}\right)=q}\mathcal{D}q\left(t\right)\,\exp\left(\frac{i}{\hbar}S_{\textrm{tot }}\left[q\left(t\right);x,\overline{x};t_{f}\right]\right)\label{eq:Forward_Propagator_Integrated}
\end{equation}
Here $A$ is a fluctuating factor that corresponds to a closed loop
path integral: 
\begin{equation}
A=\prod_{\lambda}A_{\lambda}=\prod_{\lambda}\sqrt{\frac{\omega_{\lambda}}{2\pi i\hbar\sin\left(\omega_{\lambda}t_{f}\right)}}
\end{equation}
while the action $S_{\textrm{tot}}$ is the composition of the action
of two systems, which is functionally dependent only on $q\left(t\right)$.
Explicitly:

\begin{equation}
S_{\textrm{tot }}\left[q\left(t\right);x,\overline{x};t\right]=S_{q}\left[q\left(t\right)\right]+S_{x}\left[q\left(t\right);x,\overline{x};t_{f}\right]\label{eq:Total_Action}
\end{equation}
where $S_{q}$ is the open system action
\begin{equation}
S_{q}\left[q\left(t\right)\right]=\int_{0}^{t_{f}}\textrm{d}t\ \,L_{q}\left[q\left(t\right)\right]=\int_{0}^{t_{f}}\textrm{d}t\ \,\left[\frac{1}{2}m\dot{q}^{2}\left(t\right)-V\left(q\left(t\right)\right)\right]\label{eq:Open_System_Action}
\end{equation}
and $S_{x}$ is the \emph{classical }action of a set of displaced
harmonic oscillators for an external ``force'' given by $g\left(q\left(t\right)\right)$.
This has no functional dependence on the $x$ coordinates; $S_{x}$
only depends on the limits of the path integral over the environment:

\[
S_{x}\left[q(t);x,\overline{x};t_{f}\right]=\sum_{\lambda}\left\{ \frac{\omega_{\lambda}}{\sin\left(\omega_{\lambda}t_{f}\right)}\left[\frac{1}{2}\left(x_{\lambda}^{2}+\bar{x}_{\lambda}^{2}\right)\cos\left(\omega_{\lambda}t_{f}\right)-x_{\lambda}\bar{x}_{\lambda}\right.\right.
\]

\[
+\frac{x_{\lambda}}{\omega_{\lambda}}\int_{0}^{t_{f}}\textrm{d}t\,g_{\lambda}\left(t\right)\sin\left(\omega_{\lambda}t\right)+\frac{\bar{x}_{\lambda}}{\omega_{\lambda}}\int_{0}^{t_{f}}\textrm{d}t\,g_{\lambda}\left(t\right)\sin\left(\omega_{\lambda}\left(t_{f}-t\right)\right)
\]
\begin{equation}
\left.\left.-\frac{1}{\omega_{\lambda}^{2}}\int_{0}^{t_{f}}\int_{0}^{t}\textrm{d}t\textrm{d}t^{\prime}\,g_{\lambda}\left(t\right)g_{\lambda}\left(t^{\prime}\right)\sin\left(\omega_{\lambda}\left(t_{f}-t\right)\right)\sin\left(\omega_{\lambda}t^{\prime}\right)\right]\right\} \label{eq:Environment_+_Interaction_Action}
\end{equation}

In the final equation above, we have abbreviated by setting $g\left(q\left(t\right)\right)=g\left(t\right)\equiv\left\{ g_{\lambda}(t)\right\} $,
in addition to the limits $x\left(t_{f}\right)=x\equiv\left\{ x_{\lambda}\right\} $
and $x\left(0\right)=\bar{x}\equiv\left\{ \overline{x}_{\lambda}\right\} $. 

The backward propagator has a similar expression as compared to the
forward propagator in Eq. (\ref{eq:Forward_Propagator_Integrated}):
\begin{equation}
U\left(\bar{q}^{\prime},\bar{x}^{\prime},0;q^{\prime},x,t_{f}\right)=A^{*}\int_{q^{\prime}\left(t_{f}\right)=q^{\prime}}^{q^{\prime}\left(0\right)=\overline{q}^{\prime}}\mathcal{D}q^{\prime}\left(t\right)\,\exp\left(-\frac{i}{\hbar}S_{\textrm{tot }}\left[q^{\prime}\left(t\right);x,\overline{x}^{\prime};t_{f}\right]\right)\label{eq:Backward_Propagator_Integrated}
\end{equation}
with the same expression (\ref{eq:Total_Action}) for the action,
but using the substitution $\overline{x}\rightarrow\overline{x}^{\prime}$.
The abbreviation $g\left(q^{\prime}\left(t\right)\right)=g^{\prime}\left(t\right)\equiv\left\{ g_{\lambda}^{\prime}(t)\right\} $
will also be used when referring to the backward propagator. 

As well as the propagators, the initial density matrix may also be
expressed as a path integral over both the open system and environmental
coordinates. After performing the same integration over the environment
as for the propagators, we obtain:

\begin{equation}
\rho_{\beta}\left(\bar{q},\bar{x};\bar{q}^{\prime},\bar{x}^{\prime}\right)=\frac{A^{E}}{Z_{\beta}}\int_{\bar{q}(0)=\bar{q}^{\prime}}^{\bar{q}(\hbar\beta)=\bar{q}}\mathcal{D}\bar{q}\left(\tau\right)\,\exp\left(-\frac{1}{\hbar}S_{\textrm{tot}}^{E}\left[\bar{q}(\tau);\bar{x},\bar{x}^{\prime};\hbar\beta\right]\right)\label{eq:Thermal_Density_Matrix_Path_Integral}
\end{equation}
\begin{equation}
A^{E}=\prod_{\lambda}A_{\lambda}^{E}=\prod_{\lambda}\sqrt{\frac{\omega_{\lambda}}{2\pi\hbar\sinh\left(\omega_{\lambda}\hbar\beta\right)}}
\end{equation}
Here $Z_{\beta}$ is the partition function for the total system,
while $S_{\textrm{tot}}^{E}$ is the \emph{Euclidean }action, defined
as the Wick rotation of $S_{\textrm{tot }}\left[\bar{q}(\tau);\bar{x},\bar{x}^{\prime};\hbar\beta\right]$.
Using the notation $g(\bar{q}(t))=\bar{g}(t)\equiv\left\{ \overline{g}_{\lambda}(t)\right\} $,
$\bar{x}(\hbar\beta)=\bar{x}\equiv\left\{ \overline{x}_{\lambda}\right\} $
and $\bar{x}(0)=\bar{x}^{\prime}\equiv\left\{ \overline{x}_{\lambda}^{\prime}\right\} $,
we obtain a familiar (albeit Wick rotated) definition for $S_{\textrm{tot}}^{E}$
(see, e.g., \cite{Grabert1988}):
\[
S_{\textrm{tot}}^{E}\left[\bar{q}(\tau);\bar{x},\bar{x}^{\prime};\hbar\beta\right]=S_{q}^{E}\left[\bar{q}(\tau)\right]+S_{x}^{E}\left[\bar{q}(\tau);\bar{x},\overline{x}^{\prime};\hbar\beta\right]
\]
where the system and bath contributions are given as follows:
\begin{equation}
S_{q}^{E}\left[\bar{q}(\tau)\right]=\int_{0}^{\hbar\beta}\mathrm{d}\tau\,L_{q}^{E}\left[\bar{q}(\tau)\right]=\int_{0}^{\hbar\beta}\mathrm{d}\tau\,\left[\frac{1}{2}m\dot{\bar{q}}^{2}(\tau)+V(\bar{q}(\tau))\right]
\end{equation}
and

\[
S_{x}^{E}\left[\bar{q}(\tau);\bar{x},\overline{x}^{\prime};\hbar\beta\right]=\sum_{\lambda}\left\{ \left[\frac{\omega_{\lambda}}{\sinh\left(\omega_{\lambda}\hbar\beta\right)}\bigg[\frac{1}{2}\left(\bar{x}_{\lambda}^{2}+\bar{x}_{\lambda}^{\prime2}\right)\cosh\left(\omega_{\lambda}\hbar\beta\right)-\bar{x}_{\lambda}\bar{x}_{\lambda}^{\prime}\right.\right.
\]

\[
-\frac{\bar{x}_{\lambda}}{\omega_{\lambda}}\int_{0}^{\hbar\beta}\textrm{d}\tau\,\ \bar{g}_{\lambda}\left(\tau\right)\sinh\left(\omega_{\lambda}\tau\right)-\frac{\bar{x}_{\lambda}^{\prime}}{\omega_{\lambda}}\int_{0}^{\hbar\beta}\textrm{d}\tau\,\ \bar{g}_{\lambda}\left(\tau\right)\sinh\left(\omega_{\lambda}\left(\hbar\beta-\tau\right)\right)
\]
\begin{equation}
\left.\left.-\frac{1}{\omega_{\lambda}^{2}}\int_{0}^{\hbar\beta}\int_{0}^{\tau}\textrm{d}\tau\textrm{d}\tau^{\prime}\ \,\bar{g}_{\lambda}\left(\tau\right)\bar{g}_{\lambda}\left(\tau^{\prime}\right)\sinh\left(\omega_{\lambda}\left(\hbar\beta-\tau\right)\right)\sin\left(\omega_{\lambda}\tau^{\prime}\right)\right]\right\} 
\end{equation}

Following Ref. \cite{Grabert1988}, we now also define a new partition
function $Z=Z_{\beta}/Z_{B}$ in terms of the partition functions
of the total system $Z_{\beta}$ and the (isolated) environment 
\begin{equation}
Z_{B}=\prod_{\lambda}\frac{1}{2\sinh\left(\frac{1}{2}\omega_{\lambda}\hbar\beta\right)}
\end{equation}
After substituting the path integral and partition function expressions
into Eq. (\ref{eq:Reduced_Density_Normal_Mode_Coordinates}), we obtain
an expression for the reduced density matrix after integrating over
the environmental trajectories:

\[
\rho_{t_{f}}\left(q;q^{\prime}\right)=\frac{1}{Z}\int\mathrm{d}\bar{q}\mathrm{d}\bar{q}^{\prime}\mathcal{\mathcal{D}}q\left(t\right)\mathcal{D}\bar{q}\left(\tau\right)\mathcal{D}q^{\prime}\left(t\right)\,\mathcal{F}\left[q\left(t\right),q^{\prime}\left(t\right),\bar{q}\left(\tau\right)\right]
\]

\begin{equation}
\times\exp\left[\frac{i}{\hbar}\int_{0}^{t_{f}}\mathrm{d}t\,L_{q}\left[q\left(t\right)\right]-\frac{i}{\hbar}\int_{0}^{t_{f}}\mathrm{d}t\,L_{q}\left[q^{\prime}\left(t\right)\right]-\frac{1}{\hbar}\int_{0}^{\hbar\beta}\mathrm{d}\tau\,L_{q}^{E}\left(\bar{q}\left(\tau\right)\right)\right]\label{eq:Reduced_Density_With_Influence_Functional}
\end{equation}
The limits of the path integrals here are the same as above. The normalising
constant $Z$ in the equilibrium density operator is not generally
known, and this issue will be discussed in Section \ref{sec:The-Extended-Stochastics}.

The \emph{influence functional} $\mathcal{F}\left[q,q^{\prime},\bar{q}\right]$
contains the full path integral over the environment. It is fully
factorised over the normal modes $\lambda$, and for each mode is
composed of a product of three terms:

\begin{equation}
\mathcal{F}\left[q\left(t\right),q^{\prime}\left(t\right),\bar{q}\left(\tau\right)\right]=\frac{1}{Z_{B}}\prod_{\lambda}\int\mathrm{d}x_{\lambda}\mathrm{d}\bar{x}_{\lambda}\mathrm{d}\bar{x}_{\lambda}^{\prime}\ F_{\lambda}\left[q_{\lambda}\left(t\right),x_{\lambda},\bar{x}_{\lambda}\right]F_{\lambda}^{E}\left[\bar{q}_{\lambda}\left(\tau\right),\bar{x}_{\lambda},\bar{x}_{\lambda}^{\prime}\right]F_{\lambda}^{*}\left[q_{\lambda}^{\prime}\left(t\right),x_{\lambda},\bar{x}_{\lambda}^{\prime}\right]\label{eq:Influence_Functional_As_Product}
\end{equation}
where

\[
F_{\lambda}\left[q_{\lambda}\left(t\right),x_{\lambda},\bar{x}_{\lambda}\right]=A_{\lambda}\exp\left\{ \frac{i\omega_{\lambda}}{\hbar\sin\left(\omega_{\lambda}t_{f}\right)}\left[\frac{1}{2}\left(x_{\lambda}^{2}+\bar{x}_{\lambda}^{2}\right)\cos\left(\omega_{\lambda}t_{f}\right)-x_{\lambda}\bar{x}_{\lambda}\right.\right.
\]
\[
+\frac{x_{\lambda}}{\omega_{\lambda}}\int_{0}^{t_{f}}\textrm{d}t\,\ g_{\lambda}\left(t\right)\sin\left(\omega_{\lambda}t\right)+\frac{\bar{x}_{\lambda}}{\omega_{\lambda}}\int_{0}^{t_{f}}\textrm{d}t\,\ g_{\lambda}\left(t\right)\sin\left(\omega_{\lambda}\left(t_{f}-t\right)\right)
\]

\begin{equation}
\left.\left.-\frac{1}{\omega_{\lambda}{}^{2}}\int_{0}^{t_{f}}\int_{0}^{t}\textrm{d}t\textrm{d}t^{\prime}\,\ g_{\lambda}\left(t\right)g_{\lambda}\left(t^{\prime}\right)\sin\left(\omega_{\lambda}\left(t_{f}-t\right)\right)\sin\left(\omega_{\lambda}t^{\prime}\right)\right]\right\} 
\end{equation}

\[
F_{\lambda}^{*}\left[q_{\lambda}^{\prime}\left(t\right),x_{\lambda},\bar{x}_{\lambda}^{\prime}\right]=A_{\lambda}^{*}\exp\left\{ -\frac{i\omega_{\lambda}}{\hbar\sin\left(\omega_{\lambda}t_{f}\right)}\left[\frac{1}{2}\left(x_{\lambda}^{2}+\bar{x}{}_{\lambda}^{\prime2}\right)\cos\left(\omega_{\lambda}t_{f}\right)-x_{\lambda}\bar{x}{}_{\lambda}^{\prime}\right.\right.
\]
\[
+\frac{x_{\lambda}}{\omega_{\lambda}}\int_{0}^{t_{f}}\textrm{d}t\,\ g_{\lambda}^{\prime}\left(t\right)\sin\left(\omega_{\lambda}t\right)+\frac{\bar{x}{}_{\lambda}^{\prime}}{\omega_{\lambda}}\int_{0}^{t_{f}}\textrm{d}t\,\ g_{\lambda}^{\prime}\left(t\right)\sin\left(\omega_{\lambda}\left(t_{f}-t\right)\right)
\]
\begin{equation}
\left.\left.-\frac{1}{\omega_{\lambda}{}^{2}}\int_{0}^{t_{f}}\int_{0}^{t}\textrm{d}t\textrm{d}t'\,\ g_{\lambda}^{\prime}\left(t\right)g_{\lambda}^{\prime}\left(t^{\prime}\right)\sin\left(\omega_{\lambda}\left(t_{f}-t\right)\right)\sin\left(\omega_{\lambda}t^{\prime}\right)\right]\right\} 
\end{equation}
 
\[
F_{\lambda}^{E}\left[\bar{q}_{\lambda}\left(\tau\right),\bar{x}_{\lambda},\bar{x}_{\lambda}^{\prime}\right]=A_{\lambda}^{E}\exp\left\{ -\frac{\omega_{\lambda}}{\hbar\sinh\left(\omega_{\lambda}\hbar\beta\right)}\left[\frac{1}{2}\left(\bar{x}_{\lambda}^{2}+\bar{x}{}_{\lambda}^{\prime2}\right)\cosh\left(\omega_{\lambda}\hbar\beta\right)-\bar{x}_{\lambda}\bar{x}_{\lambda}^{\prime}\right.\right.
\]
\[
-\frac{\bar{x}_{\lambda}}{\omega_{\lambda}}\int_{0}^{\hbar\beta}\textrm{d}\tau\,\ \bar{g}_{\lambda}\left(\tau\right)\sinh\left(\omega_{\lambda}\tau\right)-\frac{\bar{x}_{\lambda}^{\prime}}{\omega_{\lambda}}\int_{0}^{\hbar\beta}\textrm{d}\tau\,\ \bar{g}_{\lambda}\left(\tau\right)\sinh\left(\omega_{\lambda}\left(\hbar\beta-\tau\right)\right)
\]

\begin{equation}
\left.\left.-\frac{1}{\omega_{\lambda}^{2}}\int_{0}^{\hbar\beta}\int_{0}^{\tau}\textrm{d}\tau\textrm{d}\tau^{\prime}\,\ \bar{g}_{\lambda}\left(\tau\right)\bar{g}_{\lambda}\left(\tau^{\prime}\right)\sinh\left(\omega_{\lambda}\left(\hbar\beta-\tau\right)\right)\sin\left(\omega_{\lambda}\tau^{\prime}\right)\right]\right\} 
\end{equation}

In order to calculate the influence functional, we notice that the
calculation can be performed for each mode $\lambda$ separately.
Then, the integrand in the triple integral over $x_{\lambda}$, $\overline{x}_{\lambda}$
and $\overline{x}_{\lambda}^{\prime}$ contains an exponential function
with a quadratic polynomial over these variables, and is hence a Gaussian.
This can therefore be directly integrated. We first note that all
pre-exponential factors in the influence functional after the integration
multiply exactly to one. Indeed, the introduction of the partition
function of the environment $Z_{B}$ in Eq. (\ref{eq:Influence_Functional_As_Product})
is to ensure that in the case of no interactions between the system
and the environment, the influence functional $\mathcal{F}\left[q,q^{\prime},\bar{q}\right]$
is unity. If $P_{\lambda}$ is the pre-exponential factor appearing
after the triple integration over $x_{\lambda}$, $\overline{x}_{\lambda}$
and $\overline{x}_{\lambda}^{\prime}$ in Eq. (\ref{eq:Influence_Functional_As_Product})
for one mode, then the overall exponential prefactor $J$ for the
influence functional after some simple algebra is one:

\begin{equation}
J=\frac{AA^{*}A^{E}}{Z_{B}}\prod_{\lambda}P_{\lambda}=1
\end{equation}

After performing the complete integration of Eq. (\ref{eq:Influence_Functional_As_Product}),
we find the following exponential expression for the influence functional
(cf. \cite{Feynman-Vernon-1963,Grabert1988}):

\begin{equation}
\mathcal{F}\left[q,q',\bar{q}\right]=\exp\left(-\frac{1}{\hbar}\Phi\left[q,q^{\prime},\bar{q}\right]\right)\equiv\exp\left(-\frac{1}{\hbar}\sum_{\lambda}\Phi_{\lambda}\left[q,q^{\prime},\bar{q}\right]\right)\label{eq:Final_Influence_Functional}
\end{equation}
where $\Phi=\sum_{\lambda}\Phi_{\lambda}$ is the \emph{influence
phase:}

\[
\Phi_{\lambda}\left[q,q^{\prime},\bar{q}\right]=-\int_{0}^{\hbar\beta}\mathrm{d}\tau\int_{0}^{\tau}\mathrm{d}\tau^{\prime}\,K_{\lambda}\left(i\tau^{\prime}-i\tau\right)\bar{g}_{\lambda}\left(\tau\right)\bar{g}_{\lambda}\left(\tau^{\prime}\right)-i\int_{0}^{\hbar\beta}\mathrm{d}\tau\int_{0}^{t_{f}}\mathrm{d}t\,K_{\lambda}\left(t-i\tau\right)\bar{g}_{\lambda}\left(\tau\right)\left(g_{\lambda}\left(t\right)-g_{\lambda}^{\prime}\left(t\right)\right)
\]
\begin{equation}
+\int_{0}^{t_{f}}\mathrm{d}t\int_{0}^{t}\mathrm{d}t^{\prime}\,\left(g_{\lambda}\left(t\right)-g_{\lambda}^{\prime}\left(t\right)\right)\left[K_{\lambda}\left(t-t^{\prime}\right)g_{\lambda}\left(t^{\prime}\right)-K_{\lambda}^{*}\left(t-t^{\prime}\right)g_{\lambda}^{\prime}\left(t^{\prime}\right)\right]\label{eq:Influence_Phase_K_No_Splitting}
\end{equation}
The term multiplying the various $g_{\lambda}$ within the integrals
is the kernel: 

\begin{equation}
K_{\lambda}\left(\theta\right)=\frac{\cosh\left(\omega_{\lambda}\left(\frac{\hbar\beta}{2}-i\theta\right)\right)}{2\omega_{\lambda}\sinh\left(\frac{1}{2}\beta\hbar\omega_{\lambda}\right)}\label{eq:Kernel_K}
\end{equation}
Note that the kernel appears in three forms, depending on purely imaginary
times, $K_{\lambda}\left(i\tau^{\prime}-i\tau\right)$, real times,
$K_{\lambda}\left(t-t^{\prime}\right)$, and complex times, $K_{\lambda}\left(t-i\tau\right)$.
It will be useful later in the derivation to split the kernel into
its real $K_{\lambda}^{R}$ and imaginary $K_{\lambda}^{I}$ parts.
For real times this produces,
\begin{equation}
K_{\lambda}^{R}\left(t\right)=\frac{1}{2\omega_{\lambda}}\coth\left(\frac{1}{2}\hbar\beta\omega_{\lambda}\right)\cos\left(\omega_{\lambda}t\right)\label{eq:Real_K_Real_Times}
\end{equation}
\begin{equation}
K_{\lambda}^{I}(t)=-\frac{1}{2\omega_{\lambda}}\sin\left(\omega_{\lambda}t\right)\label{eq:Imaginary_K_Real_Times}
\end{equation}
and for complex times,

\begin{equation}
K_{\lambda}^{R}\left(t-i\tau\right)=\frac{1}{2\omega_{\lambda}}\left[\coth\left(\frac{1}{2}\omega_{\lambda}\hbar\beta\right)\cosh\left(\omega_{\lambda}\tau\right)-\sinh\left(\omega_{\lambda}\tau\right)\right]\cos\left(\omega_{\lambda} t\right)\label{eq:Real_K_Complex_Times}
\end{equation}

\begin{equation}
K^{I}(t-i\tau)=-\frac{1}{2\omega_{\lambda}}\left[\cosh\left(\omega\tau\right)+\sinh\left(\omega_{\lambda}\tau\right)\coth\left(\frac{1}{2}\omega_{\lambda}\hbar\beta\right)\right]\sin\left(\omega_{\lambda} t\right)\label{eq:Imaginary_K_Complex_Times}
\end{equation}
while for purely imaginary times the kernel is real, 
\begin{equation}
K_{\lambda}(i\tau)=K_{\lambda}^{e}(\tau)+K_{\lambda}^{o}(\tau)\label{eq:Im_Time_Kernel_Split}
\end{equation}
and consisting of even and odd components: 
\begin{equation}
K_{\lambda}^{o}(\tau)=\frac{1}{2\omega_{\lambda}}\sinh\left(\omega_{\lambda}\tau\right)\label{eq:K_Imaginary_Times_Odd}
\end{equation}
\begin{equation}
K_{\lambda}^{e}(\tau)=\frac{1}{2\omega_{\lambda}}\cosh\left(\omega_{\lambda}\tau\right)\coth\left(\frac{1}{2}\omega_{\lambda}\hbar\beta\right)\label{eq:K_Imaginary_Times_Even}
\end{equation}
If for real times we also define new sum and difference interaction
functions \cite{Kleinert1995}, 
\begin{equation}
\epsilon_{\lambda}\left(t\right)=g_{\lambda}\left(t\right)-g_{\lambda}^{\prime}\left(t\right)\quad\mbox{and}\quad y_{\lambda}(t)=\frac{1}{2}\left(g_{\lambda}\left(t\right)+g_{\lambda}^{\prime}\left(t\right)\right)\label{eq:New_Coords}
\end{equation}
and substitute these expressions into Eq. (\ref{eq:Influence_Phase_K_No_Splitting}),
the single mode influence phase can now be expressed as:

\begin{gather*}
\Phi_{\lambda}\left[q,q^{\prime},\bar{q}\right]=-\int_{0}^{\hbar\beta}\mathrm{d}\tau\int_{0}^{\hbar\beta}\mathrm{d}\tau^{\prime}\,\frac{1}{2}\left[K_{\lambda}^{e}\left(\tau'-\tau\right)-K_{\lambda}^{o}\left(\left|\tau'-\tau\right|\right)\right]\bar{g}_{\lambda}\left(\tau\right)\bar{g}_{\lambda}\left(\tau^{\prime}\right)
\end{gather*}
\[
-i\int_{0}^{\hbar\beta}\mathrm{d}\tau\int_{0}^{t_{f}}\mathrm{d}t\,\left[K_{\lambda}^{R}\left(t-i\tau\right)+K_{\lambda}^{I}\left(t-i\tau\right)\right]\bar{g}_{\lambda}\left(\tau\right)\epsilon_{\lambda}\left(t\right)
\]

\begin{equation}
+\frac{1}{2}\int_{0}^{t_{f}}\mathrm{d}t\int_{0}^{t_{f}}\mathrm{d}t^{\prime}\,K_{\lambda}^{R}\left(t-t^{\prime}\right)\epsilon_{\lambda}\left(t\right)\epsilon_{\lambda}\left(t^{\prime}\right)+2i\int_{0}^{t_{f}}\mathrm{d}t\int_{0}^{t_{f}}\mathrm{d}t^{\prime}\,\left[\theta\left(t-t^{\prime}\right)K_{\lambda}^{I}\left(t-t^{\prime}\right)\right]\epsilon_{\lambda}\left(t\right)y_{\lambda}\left(t^{\prime}\right)\label{eq:Influence_Phase_K_Real_Imaginary_Split}
\end{equation}
The final two terms in this expression are a generalisation of the
well known Feynman-Vernon influence functional \cite{Feynman-Vernon-1963},
with the remaining terms arising from the incorporation of a non-partitioned
initial density matrix. Note that, compared to Eq. (\ref{eq:Influence_Phase_K_No_Splitting}),
the above expression was modified to ensure identical limits in the
double integrals over the times $t,t^{\prime}$ and $\tau,\tau^{\prime}$. 

The influence phase still contains the normal mode interaction term
$g_{\lambda}$. Using Eq. (\ref{eq:Normal_Mode_Interaction_Term}),
we can re-express the phase in terms of the original interaction given
in the site representation. The normal mode transformation did not
change the $q$ coordinates themselves, so there is no difference
between representations in the path integral measure or action $S_{q}$
in Eq. (\ref{eq:Reduced_Density_With_Influence_Functional}). The
system-bath interaction term contained in the influence functional
\emph{will }have a different form however, and hence the influence
phase has a non-trivial alternative representation in terms of functions
$f_{i}(t)\equiv f_{i}\left(q\left(t\right)\right)$ rather than $g_{\lambda}\left(q\left(t\right)\right)$.
In this representation the sum and difference functions 

\begin{equation}
v_{i}\left(t\right)=f_{i}\left(t\right)-f_{i}^{\prime}\left(t\right)\quad\mbox{and}\quad r_{i}\left(t\right)=\frac{1}{2}\left(f_{i}\left(t\right)+f{}_{i}^{\prime}\left(t\right)\right)\label{eq:Sum_Difference_Forces}
\end{equation}
can conveniently be introduced, using $f_{i}^{\prime}(t)\equiv f_{i}\left(q^{\prime}\left(t\right)\right)$.
Substituting Eq. (\ref{eq:Normal_Mode_Interaction_Term}) into these,
we can relate the sum and difference functions (\ref{eq:New_Coords})
between the normal mode and site representations:

\begin{equation}
v_{i}\left(t\right)=\frac{1}{\sqrt{m_{i}}}\sum_{\lambda}e_{\lambda i}\epsilon_{\lambda}\left(t\right)\ \ \mbox{and}\ \ r_{i}\left(t\right)=\frac{1}{\sqrt{m_{i}}}\sum_{\lambda}e_{\lambda i}y_{\lambda}\left(t\right)
\end{equation}

The influence phase in the site representation is most easily expresed
by defining new kernels from those derived using normal modes
\begin{equation}
L_{ij}^{R,I}\left(t\right)=\frac{1}{\sqrt{m_{i}m_{j}}}\sum_{\lambda}e_{\lambda i}e_{\lambda j}K_{\lambda}^{R,I}\left(t\right)\label{eq:Real_L_all_times}
\end{equation}
\begin{equation}
L_{ij}\left(t-i\tau\right)=\frac{1}{\sqrt{m_{i}m_{j}}}\sum_{\lambda}e_{\lambda i}e_{\lambda j}K_{\lambda}\left(t-i\tau\right)\label{eq:L_Complex_Time}
\end{equation}

\begin{equation}
L_{ij}^{e}\left(\tau\right)=\frac{1}{\sqrt{m_{i}m_{j}}}\sum_{\lambda}\frac{e_{\lambda i}e_{\lambda j}}{2\omega_{\lambda}}\coth\left(\frac{1}{2}\hbar\beta\omega_{\lambda}\right)\cosh\left(\omega_{\lambda}\tau\right)\label{eq:L_Even}
\end{equation}

\begin{equation}
L_{ij}^{o}\left(\tau\right)=\frac{1}{\sqrt{m_{i}m_{j}}}\sum_{\lambda}\frac{e_{\lambda i}e_{\lambda j}}{2\omega_{\lambda}}\sinh\left(\omega_{\lambda}\tau\right)\label{eq:L_Odd}
\end{equation}
 so that the influence phase can be re-expressed in terms of the site
interactions:

\begin{equation}
\Phi\left[q,q^{\prime},\bar{q}\right]=\sum_{ij}\Phi_{ij}\left[q,q^{\prime},\bar{q}\right]\label{eq:Site_Rep._Influence_Phase}
\end{equation}

\[
\Phi_{ij}\left[q,q^{\prime},\bar{q}\right]=-\int_{0}^{\hbar\beta}\mathrm{d}\tau\int_{0}^{\hbar\beta}\mathrm{d}\tau^{\prime}\,\frac{1}{2}\bar{f}_{i}(\tau)\left[L_{ij}^{e}\left(\tau'-\tau\right)-L_{ij}^{o}\left(\left|\tau'-\tau\right|\right)\right]\bar{f}_{j}\left(\tau^{\prime}\right)
\]

\[
-i\int_{0}^{\hbar\beta}\mathrm{d}\tau\int_{0}^{t_{f}}\mathrm{d}t\,v_{i}\left(t\right)L_{ij}\left(t-i\tau\right)\bar{f}_{j}\left(\tau\right)
\]

\begin{gather}
+\frac{1}{2}\int_{0}^{t_{f}}\mathrm{d}t\int_{0}^{t_{f}}\mathrm{d}t^{\prime}\,v_{i}\left(t\right)L_{ij}^{R}\left(t-t^{\prime}\right)v_{j}\left(t^{\prime}\right)+2i\int_{0}^{t_{f}}\mathrm{d}t\int_{0}^{t_{f}}\mathrm{d}t^{\prime}\,v_{i}\left(t\right)\left[\theta\left(t-t^{\prime}\right)L_{ij}^{I}\left(t-t^{\prime}\right)\right]r_{j}\left(t^{\prime}\right)\label{eq:Influence_Phase_Final}
\end{gather}
where an obvious short-hand notation $f(\bar{q}(\tau))\equiv\bar{f}_{i}(\tau)$
has also been introduced. 

The influence phase expressed here contains additional complexity
compared to one derived using a standard CL model (which does not
require a normal mode transformation) \cite{Grabert1988}. After allowing
the environment to contain internal couplings, we find that the effect
of this generalisation on the form of the influence phase is not trivial:
instead of a single sum over the bath lattice in the CL model, we
have double sums in Eq. (\ref{eq:Site_Rep._Influence_Phase}), and
this will have a profound effect on the dimensionality of the stochastic
field to be introduced below.

In principle, having found the influence phase, Eq. (\ref{eq:Reduced_Density_With_Influence_Functional})
can be used to describe the exact dynamics of the open system at all
times. Path integrals are however awkward to evaluate outside of certain
special cases. The goal now is to use Eq. (\ref{eq:Reduced_Density_With_Influence_Functional})
to derive an operator expression, and hence a Liouville-von Neumann
type equation for the reduced density matrix instead. Unfortunately
the influence phase contains double integrals in two time variables
($t$ and $\tau$), meaning there is no simple method to construct
a differential equation directly out of Eq. (\ref{eq:Reduced_Density_With_Influence_Functional}).
Here we will follow previous work \cite{Schmid1982,Kleinert1995,Tsusaka1999,Stockburger2004},
and use a transformation to convert this non-local system into a local
one exactly, at the cost of introducing stochastic variables.

\section{The Two-Time Hubbard-Stratonovich transformation \label{sec:The-Two-Time-Hubbard-Stratonovich}}

In order to progress, we will use a statistical technique known as
the Hubbard-Stratonovich (HS) transformation \cite{HSTransform}.
We shall consider the most general form of such a transformation based
on a complex multivariate Gaussian distribution (cf. \cite{Stockburger2004}). 

Consider a Gaussian distribution over $N$ complex random variables
(``noises''), $z^{1}\equiv\left\{ \eta_{i}\right\} $, and their
$N$ complex conjugates, $z^{2}\equiv\left\{ \eta_{i}^{*}\right\} $:

\begin{equation}
W\left[\eta_{1},\eta_{1}^{*},...,\eta_{N},\eta_{N}^{*}\right]=\frac{\left(2\pi\right)^{-N}}{\sqrt{\det\varSigma}}\,\exp\left[-\frac{1}{2}z^{T}\Sigma z\right]\label{eq:General_Complex_Gaussian}
\end{equation}
where

\begin{equation}
z^{\alpha}=\left(\begin{array}{c}
z_{1}^{\alpha}\\
z_{2}^{\alpha}\\
\vdots\\
z_{N}^{\alpha}
\end{array}\right)
\end{equation}
is the vector of complex variables ($\alpha=1$) or their conjugate
($\alpha=2$). The total vector $z$ is therefore of size $2N$ and
is given by:

\begin{align}
z & =\left(\begin{array}{c}
z^{1}\\
z^{2}
\end{array}\right)
\end{align}
The covariance matrix $\Sigma$ can also be decomposed into a block
form 
\begin{equation}
\varSigma\equiv\left(\Sigma_{ij}^{\alpha\beta}\right)=\left(\begin{array}{cc}
\Sigma^{11} & \Sigma^{12}\\
\Sigma^{21} & \Sigma^{22}
\end{array}\right)
\end{equation}
and the correlation functions are given by the usual Gaussian identity:
\begin{equation}
\left\langle z_{i}^{\alpha}z_{j}^{\beta}\right\rangle _{z}=\left(\Sigma^{-1}\right)_{ij}^{\alpha\beta}
\end{equation}
The Fourier transform of this distribution is the complementary distribution
which can be calculated exactly: 
\begin{align}
\kappa\left[k\right] & =\int\mathrm{d}z\ W\left(z\right)\,\exp\left(iz^{T}k\right)=\exp\left(-\frac{1}{2}k^{T}\Sigma^{-1}k\right)\label{eq:Complimetary_Distribution}
\end{align}
where $k$ is a $2N$-fold vector, consisting of two size $N$ vectors
$k^{1}$ and $k^{2}$.

This equation can be interpreted as an average (with respect to the
Gaussian distribution $W$) of the exponential function, $\left\langle \exp\left(iz^{T}k\right)\right\rangle _{z}$.
Using the distribution $W$, one can also calculate the correlation
function between any two stochastic variables. Hence, the elements
of the inverse matrix $\Sigma^{-1}$ appearing in Eq. (\ref{eq:Complimetary_Distribution})
can be written via the correlation functions. The HS transformation
is essentially the relation between these two representations of the
complementary distribution:

\begin{equation}
\left\langle \exp\left(iz^{T}k\right)\right\rangle _{z}\equiv\left\langle \exp\left(i\sum_{i\alpha}z_{i}^{\alpha}k_{i}^{\alpha}\right)\right\rangle _{z}=\exp\left(-\frac{1}{2}\sum_{ij\alpha\beta}k_{i}^{\alpha}\left\langle z_{i}^{\alpha}z_{j}^{\beta}\right\rangle _{z}k_{j}^{\beta}\right)\label{eq:HS_Discrete}
\end{equation}

So far, we have considered a finite set of discrete stochastic variables
$\left\{ \eta_{i},\eta_{i}^{*}\right\} $. The preceding derivation
can be extended to (continuous) Gaussian \emph{stochastic processes}
if different stochastic variables are now associated with time instances
$t_{k}$ separated by some small time interval $\Delta$, i.e. $z_{i}^{\alpha}\to z_{i}^{\alpha}\left(t_{k}\right)$.
Here $t_{k}=k\Delta$ with $k$ running from $0$ to $n$, so that
$n\Delta=t_{f}$. Now in the limit of $\Delta\to0$, we obtain the
HS transformation for a set of continuous Gaussian stochastic processes
as follows:

\begin{equation}
\left\langle \exp\left[i\sum_{i\alpha}\int_{0}^{t_{f}}\mathrm{d}t\ \,z_{i}^{\alpha}\left(t\right)k_{i}^{\alpha}\left(t\right)\right]\right\rangle _{z(t)}=\exp\left[-\frac{1}{2}\sum_{ij\alpha\beta}\int_{0}^{t_{f}}\textrm{d}t\int_{0}^{t_{f}}\textrm{d}t^{\prime}\ k_{i}^{\alpha}\left(t\right)\left\langle z_{i}^{\alpha}\left(t\right)z_{j}^{\beta}\left(t^{\prime}\right)\right\rangle _{z(t)}k_{j}^{\beta}\left(t^{\prime}\right)\right]\label{eq:1_Time_HS_Transform_N_D}
\end{equation}
Note that integration over the noises $z(t)=\{z_{i}^{\alpha}(t)\},$
appearing in both sides of the above equation, becomes the corresponding
path integral in the continuum limit.

Using the HS transformation defined above, clear progress can be made.
Indeed, the exponent in the right hand side of Eq. (\ref{eq:1_Time_HS_Transform_N_D})
is of the same form as the Feynman-Vernon terms of the influence phase
in Eq. (\ref{eq:Influence_Phase_Final}). The correlation functions
and $k$ variables in Eq. (\ref{eq:1_Time_HS_Transform_N_D}) can
be mapped to the terms appearing in the integrands of the Feynman-Vernon
influence phase. The HS transformation can therefore be used to equate
a deterministic non-local integral exponent to a local phase involving
auxiliary stochastic terms, that must be averaged over the distribution
$W$. In a more physical sense, we can also consider the HS transformation
as converting a system of two body potentials into a set of independent
particles in a fluctuating field. The difficulty using this transformation
is that Eq. (\ref{eq:Influence_Phase_Final}) contains two time dimensions
- one real and one imaginary, with one term involving an integration
over both dimensions - requiring a generalisation of the transformation.

When we consider how the HS transformation is derived, continuous
processes and multiple variables are incorporated through the addition
of extra indices, partitioning the arbitrary sum of random complex
variables. The same procedure can be applied to introduce different
time dimensions. Starting from a discrete representation, we introduce
two sets of times, $\left\{ t_{k},k=0,\ldots,M\right\} $ and $\left\{ \tau_{k},k=0,\ldots,M^{\prime}\right\} $,
so that the exponent on the left hand side of the HS transformation
(\ref{eq:HS_Discrete}) has the form 
\begin{equation}
z^{T}k\:\Rightarrow\:\sum_{\alpha}\left(\sum_{ik}z_{i}^{\alpha}\left(t_{k}\right)k_{i}^{\alpha}\left(t_{k}\right)+\sum_{ik}\overline{z}_{i}^{\alpha}\left(\tau_{k}\right)\overline{k}_{i}^{\alpha}\left(\tau_{k}\right)\right)
\end{equation}
where we assign $t_{M}=t_{f}$ and $\tau_{M^{\prime}}=\hbar\beta$,
and we place a bar above quantities associated with the second set
of times (denoted with the real time $\tau_{k}$). Note that the number
of stochastic variables in each set (as counted by the index $i$
for the given time index $k$) may be different for barred and unbarred
fields. In the continuum limit $M,M^{\prime}\to\infty$ we obtain
for the left hand side of the HS transformation:

\begin{equation}
\kappa\left[k\left(t\right),\bar{k}\left(t\right)\right]=\left\langle \exp\left[i\sum_{\alpha i}\int_{0}^{t_{f}}\mathrm{d}t\,z_{i}^{\alpha}\left(t\right)k_{i}^{\alpha}\left(t\right)+i\sum_{\alpha i}\int_{0}^{\hbar\beta}\mathrm{d}\tau\,\bar{z}_{i}^{\alpha}\left(\tau\right)\bar{k}_{i}^{\alpha}\left(\tau\right)\right]\right\rangle _{\left\{ z(t),\bar{z}(\tau)\right\} }
\end{equation}
Correspondingly, the exponent on the right hand side of Eq. (\ref{eq:HS_Discrete})
(after the time labels are introduced), in the continuous limit becomes:
\[
\kappa\left[k\left(t\right),\bar{k}\left(\tau\right)\right]=\exp\left\{ -\frac{1}{2}\sum_{\alpha\beta ij}\left(\int_{0}^{t_{f}}\textrm{d}t\int_{0}^{t_{f}}\textrm{d}t^{\prime}\,k_{i}^{\alpha}\left(t\right)^{T}A_{ij}^{\alpha\beta}\left(t,t^{\prime}\right)k_{j}^{\beta}\left(t^{\prime}\right)\right.\right.
\]
\begin{equation}
\left.\left.+\int_{0}^{\hbar\beta}\textrm{d}\tau\int_{0}^{\hbar\beta}\textrm{d}\tau^{\prime}\,\bar{k}_{i}^{\alpha}\left(\tau\right)^{T}A_{ij}^{\alpha\beta}\left(\tau,\tau^{\prime}\right)\bar{k}_{j}^{\beta}\left(\tau^{\prime}\right)+2\int_{0}^{t_{f}}\textrm{d}t\int_{0}^{\hbar\beta}\textrm{d}\tau\,k_{i}^{\alpha}\left(t\right)^{T}A_{ij}^{\alpha\beta}\left(t,\tau\right)\bar{k}_{j}^{\beta}\left(\tau\right)\right)\right\} 
\end{equation}
where, because of the three possible combinations of times, we introduce
three types of correlation functions: 
\begin{equation}
A_{ij}^{\alpha\beta}\left(t,t^{\prime}\right)=\left\langle z_{i}^{\alpha}\left(t\right)z_{j}^{\beta}\left(t^{\prime}\right)\right\rangle _{\left\{ z(t),\bar{z}(\tau)\right\} }
\end{equation}

\begin{equation}
A_{ij}^{\alpha\beta}\left(\tau,\tau^{\prime}\right)=\left\langle \bar{z}_{i}^{\alpha}\left(\tau\right)\bar{z}_{j}^{\beta}\left(\tau^{\prime}\right)\right\rangle _{\left\{ z(t),\bar{z}(\tau)\right\} }
\end{equation}
\begin{equation}
A_{ij}^{\alpha\beta}\left(t,\tau\right)=\left\langle z_{i}^{\alpha}\left(t\right)\bar{z}_{j}^{\beta}\left(\tau\right)\right\rangle _{\left\{ z(t),\bar{z}(\tau)\right\} }
\end{equation}
In the full multivariate form, the two-time transformation is therefore
given by:

\[
\left\langle \exp\left[i\sum_{i\alpha}\left(\int_{0}^{t_{f}}\textrm{d}t\ z_{i}^{\alpha}\left(t\right)k_{i}^{\alpha}\left(t\right)+\int_{0}^{\hbar\beta}\textrm{d}\tau\ \bar{z_{i}}^{\alpha}\left(\tau\right)\bar{k}_{i}^{\alpha}\left(\tau\right)\right)\right]\right\rangle _{\left\{ z(t),\bar{z}(\tau)\right\} }
\]
\[
=\exp\left[-\frac{1}{2}\sum_{ij\alpha\beta}\left(\int_{0}^{t_{f}}\mathrm{\textrm{d}}t\int_{0}^{t_{f}}\textrm{d}t^{\prime}\,k_{i}^{\alpha}\left(t\right)A_{ij}^{\alpha\beta}\left(t,t^{\prime}\right)k_{j}^{\beta}\left(t^{\prime}\right)\right.\right.
\]

\begin{gather}
+\left.\left.\int_{0}^{\hbar\beta}\mathrm{d}\tau\int_{0}^{\hbar\beta}\textrm{d}\tau^{\prime}\,\bar{k}_{i}^{\alpha}\left(\tau\right)A_{ij}^{\alpha\beta}\left(\tau,\tau^{\prime}\right)\bar{k}_{j}^{\beta}\left(\tau^{\prime}\right)+2\int_{0}^{t_{f}}\mbox{d}t\int_{0}^{\hbar\beta}\textrm{d}\tau\,k_{i}^{\alpha}\left(t\right)A_{ij}^{\alpha\beta}\left(t,\tau\right)\bar{k}_{j}^{\beta}\left(\tau\right)\right)\right]\label{eq:HS_Final}
\end{gather}

The connection between the influence phase and the two-time Hubbard-Stratonovich
transformation should now be transparent. Notice that here in the
exponential all time integrals have either $t_{f}$ or $\hbar\beta$
as their upper limits, exactly as in the influence functional expression
(\ref{eq:Influence_Phase_Final}) for the phase. The choice for the
second time dimension to run up to $\hbar\beta$ has been made to
highlight the closeness between the influence phase in Eq. (\ref{eq:Influence_Phase_Final})
and the two-time HS transformation presented here. 

Now we would like to apply the HS transformation to the influence
functional expression given by Eqs. (\ref{eq:Final_Influence_Functional}),
(\ref{eq:Site_Rep._Influence_Phase}) and (\ref{eq:Influence_Phase_Final}).
It is clear from the structure of the exponent in the influence functional
in Eq. (\ref{eq:Influence_Phase_Final}), that auxiliary stochastic
fields should be introduced separately for each lattice site index
$i$. Moreover, there should be two pairs of the stochastic processes
for the set associated with the real time $t$,
\begin{align}
z_{i}\left(t\right) & \;\Rightarrow\;\left(\begin{array}{c}
\eta_{i}\left(t\right)\\
\eta_{i}^{*}\left(t\right)\\
\nu_{i}\left(t\right)\\
\nu_{i}^{*}\left(t\right)
\end{array}\right)
\end{align}
and one such set for the imaginary time $i\tau$:
\begin{equation}
\bar{z}_{i}\left(\tau\right)\;\Rightarrow\;\left(\begin{array}{c}
\bar{\mu}_{i}\left(\tau\right)\\
\bar{\mu}_{i}^{*}\left(\tau\right)
\end{array}\right)
\end{equation}
where we have redefined the size $M$ (number of environmental oscillators)
complex vector $z\equiv\left\{ z_{i}\right\} $ to include two noises
and their conjugates. Next, we make the following correspondence between
the functions $k_{i}(t)$ in the HS transformation (\ref{eq:HS_Final})
and the functions $v_{i}(t)$, $r_{i}(t)$ and $\overline{f}_{i}(\tau)$
appearing in the phase, Eq. (\ref{eq:Influence_Phase_Final}):
\begin{equation}
k_{i}(t)\;\Rightarrow\;\left(\begin{array}{c}
v_{i}(t)/\hbar\\
0\\
r_{i}(t)\\
0
\end{array}\right)
\end{equation}
and 
\begin{equation}
\overline{k}_{i}(\tau)\;\Rightarrow\;\left(\begin{array}{c}
i\overline{f}_{i}(\tau)/\hbar\\
0
\end{array}\right)
\end{equation}
The three pairs of stochastic processes we have introduced must ensure
that the influence functional given by Eqs. (\ref{eq:Final_Influence_Functional}),
(\ref{eq:Site_Rep._Influence_Phase}) and (\ref{eq:Influence_Phase_Final})
coincides exactly with the right hand side of the HS transformation
(\ref{eq:HS_Final}). Therefore, comparing the exponents in the right
hand side of Eq. (\ref{eq:HS_Final}) and Eq. (\ref{eq:Influence_Phase_Final}),
explicit formulas can be established for the correlation functions
$A_{ij}^{\alpha\beta}$ between the noises. These are:
\begin{equation}
\left\langle \eta_{i}\left(t\right)\eta_{j}\left(t^{\prime}\right)\right\rangle _{\left\{ z(t),\bar{z}(\tau)\right\} }=\hbar L_{ij}^{R}\left(t-t^{\prime}\right)\label{eq:Corr_F_Eta_Eta}
\end{equation}
\begin{equation}
\left\langle \eta_{i}(t)\nu_{j}\left(t^{\prime}\right)\right\rangle _{\left\{ z(t),\bar{z}(\tau)\right\} }=2i\Theta\left(t-t^{\prime}\right)L_{ij}^{I}\left(t-t^{\prime}\right)\label{eq:Corr_F_Eta_Nu}
\end{equation}

\begin{align}
\left\langle \eta_{i}\left(t\right)\bar{\mu}_{j}\left(\tau\right)\right\rangle _{\left\{ z(t),\bar{z}(\tau)\right\} } & =-\hbar\left[L_{ij}^{R}\left(t-i\tau\right)+iL_{ij}^{I}\left(t-i\tau\right)\right]\label{eq:Corr_F_Eta_Mu}
\end{align}
\begin{equation}
\left\langle \bar{\mu}_{i}\left(\tau\right)\bar{\mu}_{j}\left(\tau^{\prime}\right)\right\rangle _{\left\{ z(t),\bar{z}(\tau)\right\} }=\hbar\left[L_{ij}^{e}\left(\tau-\tau^{\prime}\right)-L_{ij}^{o}\left(\left|\tau-\tau^{\prime}\right|\right)\right]\label{eq:Corr_F_Mu_Mu}
\end{equation}

\begin{equation}
\left\langle \nu_{i}\left(t\right)\nu_{j}\left(t^{\prime}\right)\right\rangle _{\left\{ z(t),\bar{z}(\tau)\right\} }=\left\langle \nu_{i}\left(t\right)\bar{\mu}_{j}\left(\tau\right)\right\rangle _{\left\{ z(t),\bar{z}(\tau)\right\} }=0\label{eq:Corr_F_Zero}
\end{equation}
Note that the correlation functions (\ref{eq:Corr_F_Eta_Eta}) and
(\ref{eq:Corr_F_Mu_Mu}) are to be symmetric functions with respect
to the permutation $i,t\leftrightarrow j,t^{\prime}$ and $i,\tau\leftrightarrow j,\tau^{\prime}$,
respectively, and the corresponding functions $L_{ij}^{R}$ and $L_{ij}^{o,e}$
provide exactly this. 

Taking the above results and applying them to Eq. (\ref{eq:Influence_Phase_Final}),
we find that the influence functional can be described as an average
over multivariate complex Gaussian processes as follows:

\begin{equation}
\mathcal{F}\left[q,q^{\prime},\bar{q}\right]=\left\langle \exp\left[\frac{i}{\hbar}\sum_{i}\left(\int_{0}^{t_{f}}\mathrm{d}t\,\left[\eta_{i}\left(t\right)v_{i}\left(t\right)+\hbar\nu_{i}\left(t\right)r_{i}\left(t\right)\right]+i\int_{0}^{\hbar\beta}\mathrm{d}\tau\ \,\bar{\mu}_{i}\left(\tau\right)\bar{f}_{i}\left(\tau\right)\right)\right]\right\rangle _{\left\{ z\left(t\right),\overline{z}\left(\tau\right)\right\} }\label{eq:Influence_Phase_Final_Form}
\end{equation}
where the averaging is made over three pairs of complex noises (or,
equivalently, over six real noises) per lattice site of the environment. 

Importantly, the two-time HS transformation is a purely formal one,
and we are free to stipulate that the noises are pure C-numbers; this
enables us to avoid the complication of operator-valued noises. Promoting
noises to operators has been previously shown to have no effect on
the final result, as shown in Ref. \cite{Kleinert1995,Tsusaka1999}. 

Finally it is worth mentioning that the influence phase given above
does \emph{not} uniquely define the Gaussian processes that the influence
functional is averaged over after performing the mapping. The influence
phase viewed as the right hand side of the HS transformation does
not involve every possible correlation defined under the Gaussian
distribution. In particular, the conditions we impose on some correlation
functions to map the physics to the auxiliary noises do not constrain
the correlations between the complex conjugate noises, e.g. $\left\langle \eta_{i}^{*}\left(t\right)\eta_{j}^{*}\left(t^{\prime}\right)\right\rangle $.
Therefore any distribution that satisfies Eqs. (\ref{eq:Corr_F_Eta_Eta}-\ref{eq:Corr_F_Zero})
may be used in this transformation.

\section{The Extended Stochastic Liouville-von Neumann Equation\label{sec:The-Extended-Stochastics}}

Now the influence functional $\mathcal{F}\left[q,q^{\prime},\bar{q}\right]$
has been evaluated, we are able to write the expression for the reduced
density matrix in Eq. (\ref{eq:Reduced_Density_With_Influence_Functional})
explicitly. First, having introduced stochastic variables into the
equation for the density matrix, we must define a new object $\tilde{\rho}_{t}\left(q;q^{\prime}\right)$
to act as an effective, single-trajectory density matrix defined for
a particular realisation of the stochastic processes $z(t)$ and $\bar{z}(\tau)$
along its path. Inserting Eq. (\ref{eq:Influence_Phase_Final_Form})
into Eqn.(\ref{eq:Reduced_Density_With_Influence_Functional}) we
obtain:
\begin{equation}
\tilde{\rho}_{t_{f}}\left(q;q^{\prime}\right)=\frac{1}{Z}\int\mathrm{d}\bar{q}\mathrm{d}\bar{q}^{\prime}\mathcal{\mathcal{D}}q(t)\mathcal{D}\bar{q}(\tau)\mathcal{D}q^{\prime}(t)\exp\left[\frac{i}{\hbar}\tilde{S}^{+}\left[q\left(t\right)\right]-\frac{i}{\hbar}\tilde{S}^{-}\left[q^{\prime}\left(t\right)\right]-\frac{1}{\hbar}\tilde{S}^{E}\left[\bar{q}\left(\tau\right)\right]\right]\label{eq:Single-Trajectory_Density_Matrix_Path_Integral}
\end{equation}
so that the exact reduced density matrix is recovered as an average
over all noises:

\begin{equation}
\rho_{t_{f}}\left(q;q^{\prime}\right)=\left\langle \tilde{\rho}_{t_{f}}\left(q;q^{\prime}\right)\right\rangle _{\left\{ z(t),\bar{z}\left(\tau\right)\right\} }\label{eq:Exact_Density_As_Average}
\end{equation}
Above three effective actions have been introduced:

\begin{equation}
\tilde{S}^{+}\left[q\left(t\right)\right]=\int_{0}^{t_{f}}\mathrm{d}t\,\left(L_{q}\left[q\left(t\right)\right]+\sum_{i}\left[\eta_{i}\left(t\right)+\frac{\hbar}{2}\nu_{i}\left(t\right)\right]f_{i}\left(t\right)\right)=\int_{0}^{t_{f}}\mathrm{d}t\,L^{+}\left[q\left(t\right)\right]\label{eq:Plus_Action}
\end{equation}

\begin{equation}
\tilde{S}^{-}\left[q^{\prime}\left(t\right)\right]=\int_{0}^{t_{f}}\mathrm{d}t\,\left(L_{q}\left[q^{\prime}\left(t\right)\right]+\sum_{i}\left[\eta_{i}\left(t\right)-\frac{\hbar}{2}\nu_{i}\left(t\right)\right]f_{i}\left(t\right)\right)=\int_{0}^{t_{f}}\mathrm{d}t\,L^{-}\left[q^{\prime}\left(t\right)\right]\label{eq:Minus_Action}
\end{equation}
\begin{equation}
\tilde{S}^{E}\left[\bar{q}\left(\tau\right)\right]=\int_{0}^{\hbar\beta}\mathrm{d}\tau\ \left(L_{q}^{E}\left[\bar{q}\left(\tau\right)\right]+\ \bar{\mu}_{i}\left(\tau\right)\bar{f}_{i}\left(\tau\right)\right)\label{eq:Euclidean_Action}
\end{equation}
In the definitions of the effective actions we have reinserted the
original forces $f_{i}(t)$, $f_{i}\left(t^{\prime}\right)$ and $\overline{f}_{i}(\tau)$
via Eq. (\ref{eq:Sum_Difference_Forces}). It can be seen that the
actions $\tilde{S}^{+}$ and $\tilde{S}^{-}$ correspond to two different
effective Lagrangians,
\begin{equation}
\widehat{L}^{\pm}(t)=\widehat{L}_{q}\left(t\right)+\sum_{i}\left[\eta_{i}\left(t\right)\pm\frac{\hbar}{2}\nu_{i}\left(t\right)\right]\hat{f}_{i}\left(t\right)
\end{equation}
which in turn are associated with two different effective Hamiltonians:
\begin{equation}
\widehat{H}^{\pm}(t)=\widehat{H}_{q}\left(t\right)-\sum_{i}\left[\eta_{i}\left(t\right)\pm\frac{\hbar}{2}\nu_{i}\left(t\right)\right]\hat{f}_{i}\left(t\right)\label{eq:H_Plus_Minus}
\end{equation}
As was mentioned in Section \ref{sec:The-Two-Time-Hubbard-Stratonovich},
the noises are not promoted to operators but remain as $c$-numbers. 

All three path integral coordinates have now been decoupled from each
other, and as coordinate functionals may be commuted. The density
matrix in Eq. (\ref{eq:Single-Trajectory_Density_Matrix_Path_Integral})
can therefore be expressed as:

\begin{equation}
\tilde{\rho}_{t_{f}}(q;q^{\prime})=\int\mathrm{d}\bar{q}\mathrm{d}\bar{q}^{\prime}\ U^{+}\left(q,t_{f};\bar{q},0\right)\tilde{\rho}_{0}\left(\bar{q};\bar{q}^{\prime}\right)U^{-}\left(\bar{q}^{\prime},0;q^{\prime},t_{f}\right)\equiv\left\langle q\right|\widetilde{\rho}\left(t_{f}\right)\left|q^{\prime}\right\rangle \label{eq:Stochastic_Operator_Reduced_Density_Coordinate}
\end{equation}
where 
\begin{equation}
U^{+}(q,t_{f};\bar{q},0)=\int_{q(0)=\bar{q}}^{q(t_{f})=q}\mathcal{\mathcal{D}}q\left(t\right)\ \exp\left[\frac{i}{\hbar}\tilde{S}^{+}\left[q\left(t\right)\right]\right]\equiv\left\langle q\right|\widehat{U}^{+}\left(t_{f}\right)\left|\overline{q}\right\rangle \label{eq:U_Plus_propagator}
\end{equation}

\begin{equation}
U^{-}(\bar{q}^{\prime},0;q^{\prime},t_{f})=\int_{q^{\prime}(t_{f})=q^{\prime}}^{q^{\prime}(0)=\bar{q}^{\prime}}\mathcal{\mathcal{D}}q^{\prime}\left(t\right)\ \exp\left[-\frac{i}{\hbar}\tilde{S}^{-}\left[q^{\prime}\left(t\right)\right]\right]\equiv\left\langle \overline{q}^{\prime}\right|\widehat{U}^{-}\left(t_{f}\right)\left|q^{\prime}\right\rangle \label{eq:U_Minus_Propagator}
\end{equation}

\begin{equation}
\tilde{\rho}_{0}(\bar{q};\bar{q}')=\frac{1}{Z}\int_{\bar{q}(0)=\bar{q}'}^{\bar{q}(\hbar\beta)=\bar{q}}\mathcal{D}\bar{q}\left(\tau\right)\exp\left[-\frac{1}{\hbar}\tilde{S}^{E}\left[\bar{q}\left(\tau\right)\right]\right]\equiv\left\langle \bar{q}\left|\tilde{\rho}_{0}\right|\bar{q}^{\prime}\right\rangle \label{eq:Single_Trajectory_Thermal_Matrix}
\end{equation}
Notice that the forwards propagator is \emph{not }the Hermitian conjugate
of the backwards propagator because of the obvious difference in the
their respective Hamiltonians. The consequence of this is that the
equation of motion is no longer of the Liouville form, i.e. the time
derivative of the density matrix is not solely given by the commutator
with some kind of Hamiltonian. 

Within Eqs. (\ref{eq:U_Plus_propagator}) and (\ref{eq:U_Minus_Propagator})
we have also introduced the operators 

\begin{equation}
\widehat{U}^{+}\left(t_{f}\right)=\widehat{T}\exp\left(-\frac{i}{\hbar}\int_{0}^{t_{f}}\widehat{H}^{+}(t)\mbox{d}t\right)
\end{equation}

\begin{equation}
\widehat{U}^{-}\left(t_{f}\right)=\widetilde{T}\exp\left(\frac{i}{\hbar}\int_{0}^{t_{f}}\widehat{H}^{-}(t)\mbox{d}t\right)
\end{equation}

which correspond to the forward and backward propagation performed
with the different Hamiltonians $\widehat{H}^{+}$ and $\widehat{H}^{-}$,
respectively, with the corresponding chronological $\widehat{T}$
and anti-chronological $\widetilde{T}$ time-ordering operators. It
is easy to see that the coordinate representation $\left\langle q\right|\widehat{U}^{+}\left(t_{f}\right)\left|\overline{q}\right\rangle $
and $\left\langle \overline{q}^{\prime}\right|\widehat{U}^{-}\left(t_{f}\right)\left|q^{\prime}\right\rangle $
of such operators give exactly the paths integrals in these expressions.
The propagator operators satisfy the usual equations of motion 

\begin{equation}
i\hbar\partial_{t}\widehat{U}^{+}(t)=\widehat{H}^{+}(t)\widehat{U}^{+}(t)
\end{equation}

\begin{equation}
i\hbar\partial_{t}\widehat{U}^{-}(t)=-\widehat{U}^{-}(t)\widehat{H}^{-}(t)
\end{equation}

Taking Eqs. (\ref{eq:Stochastic_Operator_Reduced_Density_Coordinate})-(\ref{eq:Single_Trajectory_Thermal_Matrix}),
the reduced single-trajectory density matrix $\widetilde{\rho}\left(t_{f}\right)$
of the open system can be written as an operator evolution:

\begin{equation}
\widetilde{\rho}(t)=\widehat{U}^{+}(t)\tilde{\rho}_{0}\widehat{U}^{-}(t)\label{eq:RDM_Time_Evolution_Final}
\end{equation}

With these definitions it is possible to generate an equation of motion
for a single-trajectory reduced density matrix by simply differentiating
the above expression with respect to time:
\[
i\hbar\partial_{t}\widetilde{\rho}\left(t\right)=\widehat{H}^{+}\left(t\right)\tilde{\rho}\left(t\right)-\tilde{\rho}\left(t\right)\widehat{H}^{-}\left(t\right)
\]

\begin{gather}
=\left[\widehat{H}_{q}\left(t\right),\tilde{\rho}\left(t\right)\right]_{-}-\sum_{i}\left(\eta_{i}\left(t\right)\left[\hat{f}_{i}\left(t\right),\tilde{\rho}\left(t\right)\right]_{-}+\frac{\hbar}{2}\nu_{i}\left(t\right)\left[\hat{f}_{i}\left(t\right),\tilde{\rho}\left(t\right)\right]_{+}\right)\label{eq:ESLN_Real_Time}
\end{gather}

This, together with an equation for $\widetilde{\rho}_{0}$, which
provides an initial condition for the reduced density operator $\widetilde{\rho}(t)$,
forms the ESLN. It bears a great deal of similarity to the equation
derived by Stockburger \cite{Stockburger2004} using the partitioned
approach, and while it may be initially surprising to see a similar
(albeit generalised) equation of motion, it seems that the partition-free
initial density matrix introduced here does not change the dynamics
it evolves under. We also note that, as was mentioned above, the obtained
equation does not have the usual Liouville form because of an extra
anti-commutator term in the right hand side. This originates from
the fact that the forward and backward propagations of the reduced
density matrix in Eq. (\ref{eq:RDM_Time_Evolution_Final}), are governed
by different Hamiltonians. We note that the same equation of motion
for the reduced density matrix can also be obtained using the method
developed by Kleinert and Shabanov in Ref. \cite{Kleinert1995}.
However, their method requires some care in choosing the correct order
of the coordinates and momenta operators. It is a definite advantage
of our method that such a problem does not arise.

All that remains is to determine the new single-trajectory initial
density matrix $\widetilde{\rho}_{0}$. This is the true initial ($t=0$)
single-trajectory reduced density matrix which is obtained from the
canonical density matrix (\ref{eq:Canonical_Density}) by tracing
out the degrees of freedom of the bath. There is already a path integral
representation for this density, Eq. (\ref{eq:Single_Trajectory_Thermal_Matrix}),
but it is unwieldy and unintuitive. Once again it is best to work
backwards to obtain the corresponding effective canonical initial
density matrix operator $\widetilde{\rho}_{0}$ with the same path
integral representation. It is easy to see, however, considering an
effective operator Hamiltonian, cf. Eq. (\ref{eq:Euclidean_Action}),
\begin{equation}
\overline{H}\left(\tau\right)=H_{q}\left(\bar{q}\right)-\sum_{i}^{M}\bar{\mu}_{i}\left(\tau\right)\bar{f}_{i}\left(\tau\right)
\end{equation}
that the path integral representation of the initial density matrix
in Eq. (\ref{eq:Single_Trajectory_Thermal_Matrix}) is formally identical
to the one for the coordinate representation of the evolution operator
when time is imaginary and $\tau$ changes between zero and $\beta\hbar$.
Therefore, the initial reduced density operator can be characterised
as a propagator through imaginary time: 
\begin{equation}
\widetilde{\rho}_{0}\equiv\left.\overline{\rho}(\tau)\right|_{\tau=\beta\hbar}\label{eq:Initial_Density_Operator}
\end{equation}
using 
\begin{equation}
\overline{\rho}(\tau)=\frac{1}{Z}\widehat{\tau}\exp\left[-\frac{1}{\hbar}\int_{0}^{\tau}\mathrm{d}\tau^{\prime}\,\overline{H}\left(\tau^{\prime}\right)\right]
\end{equation}
This has the form of a time-ordered exponent with $\widehat{\tau}$
being the corresponding chronological time-ordering operator. The
latter density operator $\overline{\rho}(\tau)$ is responsible for
the thermalisation of the open system (when $\tau\rightarrow\beta\hbar$)
and will be called the quenched initial density operator. It satisfies
the Schr{\"o}dinger-like equation of motion 
\begin{equation}
-\hbar\partial_{\tau}\overline{\rho}(\tau)=\overline{H}(\tau)\overline{\rho}(\tau)\label{eq:EoM_For_Initial_Density}
\end{equation}
with the initial condition $\overline{\rho}(\tau=0)=Z^{-1}$. The
initial density $\overline{\rho}(\tau)$ must be normalised when the
final value of $\tau\equiv\beta\hbar$ is reached, i.e. $\textrm{Tr}_{q}\left[\overline{\rho}(\beta\hbar)\right]=1$,
where the trace is taken with respect to the open system only. Therefore,
the correct initial condition for $\overline{\rho}(\tau)$ can be
fixed by providing this normalisation at the end of the imaginary
time propagation (note that $Z$, as a ratio of two partition functions,
is time independent). We also observe that essentially the same result
for the reduced equilibrium density matrix was obtained in Ref. \cite{Moix2012}. 

The Hamiltonian $H_{q}$ and the interaction operators in $\overline{H}(\tau)$
have no temperature dependence; so the temperature dependence comes
entirely from an artificial ``propagation'' of the quenched density
matrix from zero to the ``time'' $\tau=\beta\hbar$. This hard limit
relating the time to the system temperature is important, as unlike
in the real time case, the quenched density matrix may diverge as
we take $\tau\to\infty$. This is a reflection of the fact that the
path integral description of the canonical density matrix is itself
only defined for finite temperature. 

\begin{figure}[H]
\begin{centering}
\includegraphics[height=10cm]{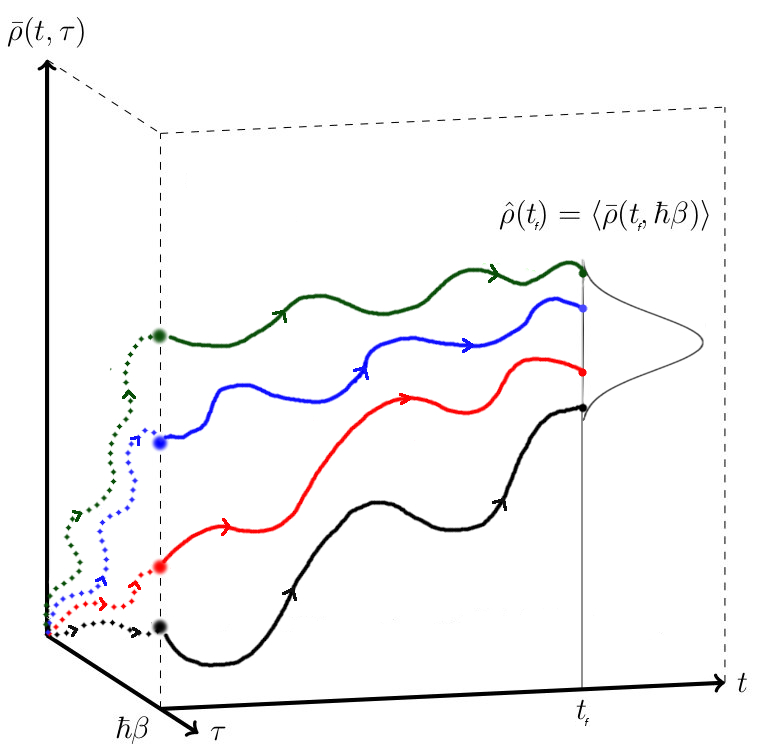}
\par\end{centering}

\caption{Representative trajectories for the evolution of the system. First
there is an evolution in imaginary time up to $\tau=\beta\hbar$,
before evolving in real time from this point up to time $t_{f}$.
Different colours correspond to different simulations associated with
particular manifestations of the noises. The average of the final
points gives the physical density matrix at that time (indicated at
time $t_{f}$).\label{fig:matrix_evolution_in_2_time_dimensions}}
\end{figure}

The equations (\ref{eq:ESLN_Real_Time}), (\ref{eq:Initial_Density_Operator})
and (\ref{eq:EoM_For_Initial_Density}) provide the complete solution
for the real time evolution of the reduced density matrix of an open
system in our partition-free approach. First of all, the initial density
matrix is obtained by propagating in imaginary time $\tau$ the quenched
density $\overline{\rho}(\tau)$ up to the final time $\tau\equiv\beta\hbar$
(the Euclidean evolution). The initial density is then normalised
which fixes the value of the partition function $Z$. Using the obtained
initial density matrix, the actual time dynamics of the reduced density
matrix $\widetilde{\rho}(t)$ are elucidated by solving Eq. (\ref{eq:ESLN_Real_Time}).
Figure \ref{fig:matrix_evolution_in_2_time_dimensions} illustrates
the evolution of trajectories through two times, as governed by the
two differential equations. First the system evolves through imaginary
time according to Eq. (\ref{eq:EoM_For_Initial_Density}) and some
realisation of the imaginary time noise trajectory $\left\{ \bar{\mu}_{i}(\tau)\right\} $.
This state then evolves through real time under Eq. (\ref{eq:ESLN_Real_Time})
using the real time noise trajectories $\left\{ \eta_{i}(t)\right\} $
and $\left\{ \nu_{i}(t)\right\} $, with the requirement that upon
averaging over realisations of these trajectories, they satisfy the
correlation functions derived in section \ref{sec:The-Two-Time-Hubbard-Stratonovich}.
The evolution along these two time dimensions is then repeated many
times using various realisations of the stochastic noises, and averaging
over many trajectories yields the physical reduced density matrix
$\widehat{\rho}\left(t\right)$ appearing in Eq. (\ref{eq:Exact_Density_As_Average}). 

\section{Discussion and conclusions \label{sec:Discussion}}

Having derived the ESLN, we should ask how it differs from previous
work. The Hamiltonian we have used is a generalisation of the Caldeira-Leggett
model, allowing for a solution in either real or frequency space.
The form of the interaction has also been generalised, but is still
limited by the essential need for an interaction to be linear in environmental
oscillator displacements. In fact, our Hamiltonian emerges naturally
from an arbitrary total system Hamiltonian by expanding atomic displacements
of the environment up to the second order. Therefore, it can be directly
applied to realistic systems. 

The fundamental result of our paper is the removal of the unphysical
partitioned initial condition which implied that the open system and
the bath were initially isolated. Following previous procedures to
accommodate a more physical partition-free approach, we applied the
special variant of the Hubbard-Stratonovich transformation that allowed
the initial condition to be determined via an auxiliary differential
equation. This allows the ESLN to make exact predictions for the transient
behaviour of the primary system when it is perturbed from equilibrium.
Additionally, when the total system is in equilibrium, the imaginary
time differential equation allows for the exact calculation of the
\emph{reduced equilibrium density matrix}. This is important, as the
stationary distribution of dissipative systems with finite couplings
has been shown to deviate from that expected under partitioned conditions
\cite{PhysRevE.84.031110}. The true distribution is described by
the ``Hamiltonian of mean force'', and Eqs. (\ref{eq:Initial_Density_Operator})
and (\ref{eq:EoM_For_Initial_Density}) provide a route to the exact
calculation of the stationary distribution. Indeed, the imaginary
time evolution has been independently derived by Moix\emph{ et al.}
\cite{Moix2012} as an exact description of an open system in interactive
equilibrium with its environment. This formulation of the equilibrium
density matrix has been used by Tanimura to develop hierarchical equations
of motion for fermionic systems \cite{Tanimura2014} under the assumption
that the environment spectral density is Ohmic. 

The ESLN represents a unification and generalisation of the differential
equations derived by Stockburger \cite{Stockburger2004} and Moix
\emph{et al.} \cite{Moix2012}, resulting in additional and highly
non-trivial constraints on the correlations between the real and imaginary
time noises. The connection between these two pieces of work was not
previously apparent, but has emerged naturally from the simultaneous
generalisation of the model Hamiltonian and the initial total density
matrix. This is the ESLN's principal advantage, and allows for a simpler
and more general closed form description of the evolution of the reduced
density matrix, as compared to hierarchical equations of motion \cite{Tanimura2014}.
We also note that our approach can easily be generalised to several
environments, e.g., for heat transport problems along similar lines
to Ref. \cite{Herve-GLE-2016}.

Extracting numerical results from the ESLN depends on the feasibility
of generating noises that satisfy the correlations outlined in section
\ref{sec:The-Two-Time-Hubbard-Stratonovich}. Real time noises of
the same type can already be efficiently calculated \cite{Stockburger2004},
and the outlook for extending this to include the imaginary time noise
is promising. Looking forward, a first application of the ESLN is
therefore likely to be a calculation of the time evolution of the
density matrix for a simple system coupled to a harmonic bath, and
the comparison between approximate partitioned and exact partition-free
methods. 

The class of problems that this model may be applied to are rather
broad. This includes a two-level spin boson system, coupled to a bath
with an arbitrary spectrum \cite{Stockburger2004}, or the heat exchange
between an arbitrary system and a bath with Ohmic dissipation \cite{Carrega2015}.
It is possible that this generalisation may also be applied to numerical
schemes for anharmonic bath models \cite{Makri1999}, and influence
functional simulations of complex systems \cite{Walters2015}. 

To summarise, the influence functional formalism has been used to
generate two stochastic differential equations that together describe
the exact evolution of an open system that begins in coupled equilibrium
with its harmonic environment. The results presented here are an extension
to existing frameworks for thermodynamic analysis in the quantum regime,
as well as offering a method for accessing the equilibrium states
of arbitrary dissipative systems.

\section*{Acknowledgements}

The authors would like to thank Prof Ulrich Weiss for his help in
verifying the Feynman-Vernon influence phase at the early stages of
this work. GMG is supported by the EPSRC Centre for Doctoral Training
in Cross-Disciplinary Approaches to Non-Equilibrium Systems (CANES,
EP/L015854/1). We also would like to thank Ian Ford and Claudia Clarke
for thought-provoking and stimulating discussions.


\end{document}